\documentclass{aa}

\usepackage{graphicx}
\usepackage{txfonts}
\newcommand{\onvire}[1]{}

\newcommand{\beq}{\begin{equation}}
\newcommand{\eeq}{\end{equation}}

\usepackage{color}
\usepackage{epsfig}
\usepackage{amssymb}
\usepackage{graphicx,natbib}
\usepackage{pdflscape}
\usepackage{longtable}
\usepackage{multirow}
\usepackage[figuresright]{rotating} 
\begin{document}

\title{The {\it Herschel}/PACS view of disks around low-mass stars in Chamaleon-I.\thanks{{\it Herschel} is an ESA space observatory with science instruments provided by European-led Principal Investigator consortia and with important participation from NASA.}}

   \author{J. Olofsson
          \inst{1}
          \and 
          L. Sz{\H u}cs
          \inst{2}
          \and
          Th. Henning
          \inst{1}
          \and 
          H. Linz
          \inst{1}
          \and 
          I. Pascucci
          \inst{3}
          \and 
          V. Joergens
          \inst{1,2}
          }

   \offprints{olofsson@mpia.de}

   \institute{Max Planck Institut f\"ur Astronomie,
     K\"onigstuhl 17, 69117 Heidelberg, Germany \\
     \email{olofsson@mpia.de}
    \and
    Universit\"at Heidelberg, Zentrum f\"ur Astronomie, Institut f\"ur Theoretische Astrophysik, Albert-Ueberle-Str. 2, 69120 Heidelberg, Germany
    \and 
    Lunar and Planetary Laboratory, The University of Arizona, Tucson, AZ 85721, USA
   }

   \date{Received \today; accepted }

 
   \abstract
   {Circumstellar disks are expected to be the birthplaces of planets. The potential for forming one or more planets of various masses is essentially driven by the initial mass of the disks, a crucial parameter for any planet formation theory. Constraining the masses of disks is a question of great interest for low-mass stars, which are expected to harbor less massive disks.}
   {We present and analyze {\it Herschel}/PACS observations of disk-bearing M-type stars that belong to the young $\sim$2\,Myr old Chamaleon-I star forming region, to better constrain the properties of the circumstellar material and the stellar mass dependance of these parameters.}
   {We used the radiative transfer code RADMC to successfully model the spectral energy distributions (SEDs) of 17 M-type stars detected at PACS wavelengths. Our modeling strategy is carefully designed so that we search for the most probable disks parameters amongst a large grid of models, via Bayesian inference, an approach that has already proven to be successful.}
   {Based on the modeling results, we first discuss the relatively low detection rates of M5 and later spectral type stars with respect to the PACS sensitivity, and argue their disks masses, or flaring indices, are likely to be low ($M_{\mathrm{disk}} \sim 10^{-5}$\,$M_\odot$, $\gamma \sim 1.1$). For M0 to M3 stars, we find a relatively broad range of disk masses ($10^{-4}$--$10^{-3}$\,$M_\odot$), scale heights, and flaring indices. Via a parametrization of dust stratification, we can reproduce the peak fluxes of the 10\,$\mu$m emission feature observed with {\it Spitzer}/IRS, and find that disks around M-type stars may display signs of dust sedimentation. We discuss a tentative correlation between the strength of the 10\,$\mu$m emission feature and the parametrized stratification.}
   {The {\it Herschel}/PACS observations of low-mass stars in Cha-I provide new constraints on their disk properties, overall suggesting that disk parameters for early M-type stars are comparable to those for more massive stars (e.g., comparable scale height and flaring angles). However, regions of the disks emitting at about 100\,$\mu$m may still be in the optically thick regime, preventing direct determination of disk masses. Thus the modeled disk masses should be considered as lower limits. Still, we are able to extend the wavelength coverage of SED models and start characterizing effects such as dust sedimentation, an effort leading the way towards ALMA observations of these low-mass stars.}
   \keywords{Stars: low-mass - circumstellar matter – Infrared: stars}
\authorrunning{Olofsson et al.}
\titlerunning{Low-mass stars in the Cha-I star forming region}

   \maketitle
%

\section{Introduction\label{sec:intro}}

Circumstellar disks are the natural by-products of the star formation process and are expected to be the birthplaces of planets. While their mass is thought to be dominated by gas with very little amount of dust (1\% in mass), it is the dust that dominates the continuum opacity of a circumstellar disk. As the disks reprocess stellar light, they can be best characterized at near- to far-infrared (IR) wavelengths, down to millimeter wavelengths. Based on the excess emission in the near-IR, the typical time-scale for the disk to dissipate is estimated to be around $\sim$\,5--6\,Myr (e.g., \citealp{Hernandez2007}). Within these few Myr, dust grains are expected to grow from $\mu$m- sized pebbles to km-sized planetesimals and eventually to planets before the disk’s material is efficiently dissipated via photo-evaporation, or grain growth and subsequent radial drift (see \citealp{Williams2011} for a recent review). An important question aims at addressing how the planet formation efficiency changes as the function of the environment, especially in the low-mass regime. Numerical simulations of planetary population synthesis, such as the ones led by \cite{Alibert2011} and \cite{Mordasini2012} have shown that the expected planet population should depend on the stellar masses and disk properties. There has been a general effort to better characterize the relationship between the stellar masses and disk properties, which can additionally be related to the formation mechanism  of very low mass stars (\citealp{Padoan2002}; \citealp{Reipurth2001}; \citealp{Bayo2011,Bayo2012}). It is therefore of prime importance to better constrain the disk characteristics, as well as dust and gas properties, around stars over a wide range of masses (e.g., \citealp{Pascucci2009}). Earlier studies based on near-IR {\it Spitzer}/IRAC and MIPS 24\,$\mu$m data (e.g. \citealp{SzHucs2010}; \citealp{Mulders2012}) investigated the vertical structure of disks as a function of the stellar mass. \citet{SzHucs2010} showed that gas disks in hydrostatic equlibrium with perfect mixing of dust and gas are inconsistent with the observations of low-mass (early M-type stars) and very low mass stars (late M dwarfs and brown dwarfs, hereafter BDs). They found that larger scale height reduction of the dust component is required for BDs than for low-mass stars when compared to the hydrostatic equilibrium solution. \cite{Mulders2012} implemented the prescription of dust settling (described in \citealp{Dullemond2004}) to hydrostatic equilibrium models and found similar turbulent mixing strengths (a parameter governing dust settling) and dust disk scale heights over a large range of stellar masses, at the same emitting regions of the disks. Supplementing near- and mid-IR observations with far-IR measurements provide additional constraints on the structure of young circumstellar disks, and the {\it Herschel} observatory (\citealp{Pilbratt2010}) has already proven successful at studying these disks (e.g., \citealp{Cieza2011}; \citealp{Joergens2012}). \cite{Harvey2012a,Harvey2012} presented the results of their {\it Herschel}/PACS Phase 1 Guaranteed Time program with which they observed a significant number of M-type stars and BDs. Thanks to a conservative Bayesian approach to model the SEDs of about 50 sources, they found that the median disk mass for their BDs sample may be smaller ($M_{\mathrm{disk}} \sim 3 \times 10^{-5}\,M_{\odot}$) than for disks around more massive T\,Tauri stars ($M_{\mathrm{disk}} \sim 10^{-3} - 10^{-1}\,M_{\odot}$). However, they found that disk properties, such as disk flaring and scale height, to be comparable with disks around T\,Tauri stars. In this paper we seek to complete these efforts by investigating disk properties in the low stellar mass regime, at the midway between T Tauri stars and brown dwarfs. We focus on stars in the M0-M5 spectral type regime in a single star forming region.

The Chamaeleon-I (Cha-I hereafter) star forming region is a perfect laboratory to study gas-rich disks around low-mass objects. Given its proximity (160-170 pc, see \citealp{Luhman2008a} for a detailed review) even M-type stars are bright enough to be observed in the far-IR with {\it Herschel}. The disk fraction in Cha-I is of about 60\%, suggesting a relatively young age ($\sim$\,2\,Myr, \citealp{Hernandez2007} and references therein, $\sim 3$\,Myr, from the isochronal age distribution, \citealp{Luhman2008a}). In this study we aim for a better characterization of the properties of disks around 17 M-type stars detected with the {\it Herschel}/PACS instruments (most likely the brightest M-type stars in Cha-I). Thanks to a careful SED modelling approach, supported with a Bayesian analysis on key parameters, we further discuss the disks properties around low-mass objects. In Sect.\,\ref{sec:obs}, we first present the {\it Herschel} observations and data processing. The modelling approach is described in Sect.\,\ref{sec:mod} and the results are detailed and discussed in Sect.\,\ref{sec:res} and \ref{sec:discuss}. We summarize our results and conclude in Sect.\,\ref{sec:conclusion}.

\section{Observations and data processing\label{sec:obs}}

In the following we first present the {\it Herschel}/PACS observations of the Cha-I star forming region, the data processing, and the sources that were successfully detected in the far-IR.

\subsection{{\it Herschel}/PACS observations}

\begin{table*}
\caption{Sources detected in the PACS observations and measured fluxes. Uncertainties on detections are 1\,$\sigma$, upper limits (denoted by ``$<$'') are 3\,$\sigma$.\label{tab:pacs}}
\begin{center}
\begin{tabular}{lccccccc}
\hline \hline
Object Name & ID & RA & Dec & Spectral type & $F_{70}$ & $F_{100}$ & $F_{160}$ \\
            &    & $[$J2000$]$ & $[$J2000$]$ &  & $[$mJy$]$ & $[$mJy$]$ & $[$mJy$]$ \\
\hline
\object{J10533978-7712338} &  1  &  10\,53\,39.78 &  $-77$\,12\,33.90 & M2.75 & $<$ 273.3 & 88.6 $\pm$ 15.7 & $<$ 162.0 \\
\object{J11044258-7741571} &  2  &  11\,04\,42.58 &  $-77$\,41\,57.13 & M4    & $<$ 231.9 & 135.7 $\pm$ 15.4 & $<$ 427.4 \\
\object{J11062554-7633418} &  3  &  11\,06\,25.55 &  $-76$\,33\,41.87 & M5.25 & $<$ 248.0 & 228.5 $\pm$ 11.8 & 284.1 $\pm$ 54.9 \\
\object{J11065906-7718535} &  4  &  11\,06\,59.07 &  $-77$\,18\,53.57 & M4.25 & $<$ 301.8 & 131.1 $\pm$ 20.2 & $<$ 461.7 \\
\object{J11071206-7632232} &  5  &  11\,07\,12.07 &  $-76$\,32\,23.23 & M0.5  & $<$ 315.3 & 295.6 $\pm$ 14.8 & 211.0 $\pm$ 58.3 \\
\object{J11074366-7739411} &  6  &  11\,07\,43.66 &  $-77$\,39\,41.15 & M0    & $<$ 615.6 & 445.0 $\pm$ 21.7 & $<$ 979.5 \\
\object{J11085464-7702129} &  7  &  11\,08\,54.64 &  $-77$\,02\,12.96 & M0.5  & $<$ 285.7 & 129.7 $\pm$ 13.4 & $<$ 159.1 \\
\object{J11091812-7630292} &  8  &  11\,09\,18.13 &  $-76$\,30\,29.25 & M1.25 & $<$ 300.9 & 120.9 $\pm$ 19.8 & $<$ 362.7 \\
\object{J11094742-7726290} &  9  &  11\,09\,47.42 &  $-77$\,26\,29.06 & M3.25 & $<$ 282.4 & 210.0 $\pm$ 14.2 & $<$ 315.4 \\
\object{J11095407-7629253} &  10 &  11\,09\,54.08 &  $-76$\,29\,25.31 & M2    & 348.3 $\pm$ 94.2 & 429.3 $\pm$ 19.8 & $<$ 373.0 \\
\object{J11095873-7737088} &  11 &  11\,09\,58.74 &  $-77$\,37\,08.88 & M1.25 & $<$ 472.5 & 226.6 $\pm$ 17.5 & $<$ 218.0 \\
\object{J11100704-7629376} &  12 &  11\,10\,07.04 &  $-76$\,29\,37.70 & M0    & $<$ 282.0 & 162.7 $\pm$ 19.7 & $<$ 352.3 \\
\object{J11104959-7717517} &  13 &  11\,10\,49.60 &  $-77$\,17\,51.70 & M2    & 709.6 $\pm$ 83.0 & 490.1 $\pm$ 14.4 & 407.8 $\pm$ 133.8 \\
\object{J11105333-7634319} &  14 &  11\,10\,53.33 &  $-76$\,34\,31.99 & M3.75 & $<$ 393.3 & 112.9 $\pm$ 19.0 & $<$ 765.8 \\
\object{J11105597-7645325} &  15 &  11\,10\,55.97 &  $-76$\,45\,32.57 & M5.75 & $<$ 294.7 & 72.6 $\pm$ 15.1 & $<$ 419.4 \\
\object{J11111083-7641574} &  16 &  11\,11\,10.83 &  $-76$\,41\,57.43 & M2.5  & $<$ 280.7 & 176.4 $\pm$ 14.9 & $<$ 445.4 \\
\object{J11120984-7634366} &  17 &  11\,12\,09.85 &  $-76$\,34\,36.61 & M5    & $<$ 287.3 & 71.6 $\pm$ 16.2 & $<$ 267.6 \\
\hline
\end{tabular}
\end{center}
\end{table*}

In this study, we make use of {\it Herschel} observations of the Cha-I region from the Guaranteed Time Key Programme led by P.\,Andr\'e at the three PACS wavelengths (70, 100 and 160\,$\mu$m, in blue, green and red filters, respectively, \citealp{Poglitsch2010}). Maps in the green and red filters were produced from OBSIDs 1342224782 and 1342224783 (PACS photometry only), both taken on the 28/07/2011 with durations of 17461 and 17721 seconds, respectively. The map in the blue filter was reduced from OBSIDs 1342213178 and 1342213179, both taken in PACS and SPIRE parallel mode, on the 22/01/2011, with durations of 12955 and 11686 seconds, respectively. Because these observations were performed in parallel mode, their sensitivity is expected to be smaller than the green map, and were therefore not used when producing the map at 160\,$\mu$m. Data processing was performed within the ``{\it Herschel} Interactive Processing Environment'' (HIPE, version 10.0.667) using build-in scripts dedicated to the data reduction of such large maps. This script makes use of highpass filtering to remove the 1/f noise and is best suited for point sources.

The target sample was drawn from previous studies of the Cha-I region by \cite{Luhman2008} and \cite{Luhman2008a}. \cite{SzHucs2010} have already compiled a list of sources in Cha-I with near-IR excess indicative of a siginificant amount of circumstellar dust, this list containing 62 M-type stars (the prime focus of this study). Here we only study the sources, that were detected at least at one of the PACS wavelengths (17 sources out of 62), but we will also discuss the detection statistics later on. The map at 100\,$\mu$m appears to be the most sensitive one compared to other wavelengths. Flux extraction was performed via aperture photometry, with aperture radii of 10$''$ for the blue and green maps, and 20$''$ for the red map. Fluxes were corrected for the aperture sizes, depending on the radii used (fluxes divided by 0.774, 0.727, and 0.800 for the blue, green, and red filters, respectively). The pixel scales for the blue and green maps are 2$''$, and 3$''$ for the red map. To estimate the uncertainties, for each sources, we randomly placed 5\,000 apertures within a circle of a radius of 150 pixels around the source position (radii of 300$''$ and 450$''$ for both pixel scales). We then build a histogram of the measured fluxes in this area and fitted a Gaussian to the distribution of fluxes. The $\sigma$ width of the Gaussian was used as the uncertainties on the measured fluxes. The center of the Gaussian fit is a good estimate of the background level, and was consequently subtracted from the source flux. These values were always smaller than the derived uncertainties. Table\,\ref{tab:pacs} summarizes the {\it Herschel}/PACS fluxes and upper limits for M-type stars detected with a signal-to-noise ratio (S/R hereafter) larger than 3 at least at one wavelength. Uncertainties on the detections are 1$\sigma$, and upper limits are 3$\sigma$. A total of 17 sources were successfully detected, with spectral types ranging from M0.5 to M5.75. Young stellar objects with later spectral types, including BDs, were not successfully detected.

\subsection{Near- and mid-IR complementary observations}

\begin{table*}
\caption{Near- and mid-IR fluxes ({\it Spitzer}/IRAC, MIPS, and WISE) from the literature for the sources detected PACS wavelengths.\label{tab:irac}}
\begin{center}
\begin{tabular}{lccccc|cccc}
\hline \hline
 & \multicolumn{5}{c|}{{\it Spitzer}/IRAC \& MIPS}  & \multicolumn{4}{c}{WISE} \\
\hline
Object Name & $F_{3.6}$ & $F_{4.5}$ & $F_{5.8}$ & $F_{8.0}$ & $F_{24}$ & $F_{3.35}$ & $F_{4.6}$ & $F_{11.56}$ & $F_{22.1}$ \\
 & $[$mJy$]$ & $[$mJy$]$ & $[$mJy$]$ & $[$mJy$]$ & $[$mJy$]$ & $[$mJy$]$ & $[$mJy$]$ & $[$mJy$]$ & $[$mJy$]$ \\
\hline
J10533978-7712338$^a$ & 7.12 & 6.58 & 6.62 & 8.00 & 55.40 & 6.40 & 6.01 & 11.14 & 59.19\\
J11044258-7741571$^b$ & 31.66 & 25.97 & 20.26 & 18.50 & 52.42 & 25.84 & 21.92 & 20.08 & 54.18\\
J11062554-7633418$^b$ & 14.21 & 12.21 & 11.61 & 11.35 & 52.42 & 12.53 & 11.89 & 13.39 & 51.88\\
J11065906-7718535$^b$ & 32.55 & 27.20 & 24.25 & 32.14 & 56.95 & 59.85 & 58.85 & 70.71 & 87.63\\
J11071206-7632232$^b$ & 97.40 & 79.90 & 71.24 & 69.03 & 115.75 & 85.33 & 69.40 & 68.03 & 133.36\\
J11074366-7739411$^b$ & 325.50 & 282.19 & 215.13 & 212.35 & 336.91 & 294.23 & 276.02 & 223.81 & 525.58\\
J11085464-7702129$^a$ & 103.88 & 104.36 & 93.05 & 84.54 & 186.86 & 99.06 & 97.58 & 96.53 & 200.19\\
J11091812-7630292$^b$ & 190.79 & 191.67 & 158.74 & 161.09 & 270.10 & 146.92 & 157.38 & 155.84 & 269.30\\
J11094742-7726290$^b$ & 68.01 & 71.54 & 51.61 & 47.76 & 121.20 & 56.90 & 57.67 & 39.11 & 117.34\\
J11095407-7629253$^b$ & 106.80 & 99.67 & 102.97 & 118.87 & 239.62 & 112.28 & 112.65 & 113.94 & 240.46\\
J11095873-7737088$^b$ & 449.32 & 451.39 & 395.09 & 397.25 & 386.83 & 399.47 & 392.05 & 344.09 & 406.11\\
J11100704-7629376$^b$ & 235.80 & 218.05 & 215.13 & 262.46 & 275.12 & 224.43 & 215.05 & 245.17 & 312.64\\
J11104959-7717517$^a$ & 75.95 & 82.14 & 90.51 & 142.91 & 452.40 & 114.36 & 114.32 & 191.90 & 448.58\\
J11105333-7634319$^b$ & 62.02 & 67.69 & 57.64 & 75.00 & 111.56 & 50.94 & 54.02 & 84.31 & 131.78\\
J11105597-7645325$^b$ & 55.53 & 52.31 & 47.94 & 46.03 & 70.39 & 49.15 & 45.35 & 35.24 & 75.90\\
J11111083-7641574$^b$ & 0.64 & 0.64 & 0.63 & 0.74 & 9.37 & 0.63 & 0.71 & 0.59 & 7.46\\
J11120984-7634366$^b$ & 51.12 & 43.91 & 35.70 & 36.56 & 63.03 & 58.82 & 51.54 & 37.11 & 60.07\\
\hline
\end{tabular}
 \begin{list}{}{}
 \item {\it Spitzer} photometry taken from: (a) \cite{Luhman2008}, (b) \cite{Luhman2008a}
 \end{list}
\end{center}
\end{table*}

In addition to the 2MASS data (\citealp{Skrutskie2006}), to construct the SEDs of the sources detected in the far-IR, we gathered photometric observations in the near- and mid-IR from literature: {\it Spitzer}/IRAC, MIPS data (\citealp{Werner2004}; \citealp{Fazio2004}; \citealp{Rieke2004}) both from \citet{Luhman2008} and \citet{Luhman2008a} and the WISE catalog (\citealp{Wright2010}). Table\,\ref{tab:irac} summarizes these fluxes for each of the source. Additionally, several stars in our final sample (15 out of 17) were observed with the IRS spectrograph (\citealp{Houck2004}) onboard {\it Spitzer}. Spectroscopic data were processed using the FEPS pipeline (see, e.g., \citealp{Bouwman2008}). 

The SEDs of the 17 M-type stars detected with PACS are displayed in Figures\,\ref{fig:SED_1} and \ref{fig:SED_2}. 2MASS and optical data are shown with filled, black circles, WISE measurements are represented with red hexagons and IRAC and MIPS observations are shown with green diamonds. PACS detections are represented by open, black circles, and upper limits by downwards black triangles. Where available, the {\it Spitzer}/IRS spectra are represented by a red line. The dashed line represents the stellar photosphere (see Sect.\,\ref{sec:mod}).

\section{SED modelling: a bayesian approach\label{sec:mod}}

\subsection{The RADMC model}

We used the radiative transfer code RADMC (\citealp{Dullemond2004}) to model the SEDs of the 17 sources detected at PACS wavelengths. A given model is first described by the stellar photosphere, and we opted for the NextGen atmospheric models (\citealp{Hauschildt1999}). Table\,\ref{tab:stellar} shows the adopted values for the effective temperature $T_{\mathrm{eff}}$, stellar luminosity ($L_{\mathrm{bol}}$) and foreground extinction ($A_{\mathrm{V}}$), taken from the literature (\citealp{Luhman2007a,Luhman2008}). We used a surface gravity of log$(g) = 3.5$, consistent with dwarf stars. For all the stars, we assumed a distance of $d_{\star} = 165$\,pc. The disk itself is then described by the inner and outer radii ($r_{\mathrm{in}}$ and $r_{\mathrm{out}}$, respectively), the disk mass ($M_{\mathrm{disk}}$, assuming a gas-to-dust ratio of 100), and a surface density profile ($\Sigma = \Sigma _0 (r / r_0)^{\alpha}$, $\alpha < 0$). The vertical distribution is assumed to follow a Gaussian profile ($\propto$ exp$[-z^2 / 2H(r)^2]$), where $H(r)$ is defined as $H(r) = H_0(r/r_0)^{\gamma}$ where $\gamma$ is the flaring exponent of the disk. The dust content is fully parametrized by the minimum and maximum grain sizes ($s_{\mathrm{min}}$ and $s_{\mathrm{max}}$), a grain size distribution (d$n(s) \propto s^p$d$s$, $p < 0$), and the optical constants of astronomical silicates (\citealp{Draine2003}) combined with the Mie scattering theory to compute the mass absorption coefficients.

\begin{table}
\caption{Stellar parameters for sources detected in the PACS observations (see Sect.\,\ref{sec:discuss} for the stellar masses determination).\label{tab:stellar}}
\begin{center}
\begin{tabular}{lcccc}
\hline \hline
Object Name & $T_\mathrm{eff}$ & $L_\mathrm{bol}$ & $A_\mathrm{V}$ & $M_\star$\\
            & $[$K$]$ & $[L_\odot]$ & & $[M_\odot]$ \\
\hline
J10533978-7712338 & 3600 & 0.032 & 2.50 & 0.64 \\
J11044258-7741571 & 3270 & 0.093 & 1.28 & 0.27 \\
J11062554-7633418 & 3091 & 0.052 & 3.59 & 0.16 \\
J11065906-7718535 & 3234 & 0.11  & 0    & 0.24 \\
J11071206-7632232 & 3778 & 0.36  & 2.21 & 0.90 \\
J11074366-7739411 & 3850 & 1.4   & 4.80 & 1.01 \\
J11085464-7702129 & 3778 & 0.34  & 3.19 & 0.90 \\
J11091812-7630292 & 3669 & 0.55  & 6.82 & 0.74 \\
J11094742-7726290 & 3379 & 0.22  & 8.02 & 0.36 \\
J11095407-7629253 & 3560 & 0.48  & 5.22 & 0.58 \\
J11095873-7737088 & 3669 & 0.84  & 1.99 & 0.74 \\
J11100704-7629376 & 3850 & 1.4   & 2.50 & 1.01 \\
J11104959-7717517 & 3560 & 0.42  & 4.15 & 0.58 \\
J11105333-7634319 & 3306 & 0.13  & 0.50 & 0.29 \\
J11105597-7645325 & 3024 & 0.13  & 0.81 & 0.12 \\
J11111083-7641574 & 3488 & 0.8$^{a}$ & 2.41 & 0.49 \\
J11120984-7634366 & 3125 & 0.15  & 0.39 & 0.18 \\
\hline
\end{tabular}
 \begin{list}{}{}
 \item (a) see Section\,\ref{sec:edge-on}.
 \end{list}
\end{center}
\end{table}

\subsection{The grid of models}

\begin{table*}
\caption{Disk parameters for the grids of models. The number in parenthesis are the number of values for each parameter.\label{tab:grid}}
\begin{center}
\begin{tabular}{lcc}
\hline \hline
Parameter  & & Values \\
\hline
$M_{\mathrm{disk}}$ & (6) & [$10^{-5}$, $5 \times 10^{-5}$, $10^{-4}$, $5 \times 10^{-4}$, $10^{-3}$, $5 \times 10^{-3}$]\,$M_{\odot}$ \\
$r_{\mathrm{in}}$ & (3) & [1, 2, 3]\,$\times r_{\mathrm{sub}}$ \\
$H_0$ & (5) & [5, 10, 15, 20, 25]\,AU \\
$H_{0,\,\mathrm{small}}$ & (4) & $H_0 +$\,[0, 1, 2, 5]\,AU \\
$\gamma$ & (7) & [1.00, 1.05, 1.10, 1.15, 1.20, 1.25, 1.30] \\
$i$ & (10) & [5, 14, 23, 32, 41, 50, 59, 68, 77, 86] \\
\hline
\end{tabular}
\end{center}
\end{table*}

Our modeling approach is motivated by the work presented in \cite{Harvey2012}. The goal is not necessarily to achieve the best possible fit to the data, given the degeneracy of certain parameters in SED modeling without spatially resolved observations. Instead, we aim at finding the most probable values for several key parameters, such as $M_{\mathrm{disk}}$ or the scale height of disks around M-type stars. We therefore decided to choose a few relevant parameters, and keep the other constant. Based on the results presented in \cite{Harvey2012}, $r_{\mathrm{out}}$ is chosen to remain constant (100\,AU) for all the objects. Determining the value of $r_{\mathrm{out}}$ is a challenging problem, even when millimeter observations are available. Only a handful of studies could constrain the outer radius of disks around low mass stars and typical values of 15--40\,AU were found (e.g., \citealp{Luhman2007b}; \citealp{Ricci2013}). Several studies investigated whether or not $r_{\mathrm{out}}$ could be inferred from SED modeling of far-IR observations (e.g., \citealp{Harvey2012a}; \citealp{Spezzi2013}) and concluded the {\it Herschel} observations mostly remained insensitive to this parameter. This could indicate that the far-IR emission arises from regions within a few AUs of the disks and that the parameter $r_{\mathrm{out}}$ mostly has a significant impact on the millimeter emission. In an effort to limit the number of free parameters we opt not to include $r_{\mathrm{out}}$ in the grid of models, an approach in line with the modeling strategies described for instance in \citet{Andrews2005,Andrews2007}; \cite{Morrow2008} and \cite{Mathews2013}. On the other end, $r_{\mathrm{in}}$ is a crucial parameter to get the near-IR shape of the SED properly. Even though our study is not focused on the inner disk structure, $r_{\mathrm{in}}$ is a free parameter in the grid. The surface density distribution has an exponent of $\alpha = -1$, a value shallower than the Minimum Mass Solar Nebula, but consistent with theoretical estimations (e.g., \citealp{Bell1997}) and observational results ($\alpha \sim -0.9$, \citealp{Andrews2009}). The grain size distribution follows a power-law with a slope in $p = -3.5$ (\citealp{Mathis1977}). The grid of models is therefore computed over the following parameters: $r_{\mathrm{in}}$, $M_{\mathrm{dust}}$, $H_0$, and $\gamma$. For $H_0$, the reference radius is set at $r_0 = 100$\,AU. Each model provides 10\,SEDs for different inclinations $i$ between 5$\degr$ and 86$\degr$ (0$\degr$ being a face-on disk).

Even though we do not aim at fitting perfectly the IRS spectra, preliminary tests have shown that the observed emission features (especially at 10\,$\mu$m) provide valuable information. We computed a first grid of models for the object J11071206-7632232, with $s_{\mathrm{min}} = 0.03$\,$\mu$m and $s_{\mathrm{max}} = 1$\,mm for the dust grains. The most probable model (see next paragraph) reproduces the overall shape of the SED but the 10\,$\mu$m emission feature was severely under-predicted by all the models. We interpreted this as a sign for dust settling, and consequently included a ``second'' population which contains only small dust grains in the upper layers of the disk, with $s_{\mathrm{min}} = 0.01$\,$\mu$m and $s_{\mathrm{max}} = 0.1$\,$\mu$m. We parametrized this second population of dust grains via another scale height reference value $H_{0,\,\mathrm{small}}$, that can be 0 (no stratification), 1, 2, or 5\,AU above the $H_0$ for the larger grains, therefore mimicking a sedimentation of the larger grains. The small grain component was set to account for 10\% of the total dust mass. The values for $s_{\mathrm{min}}$ and $s_{\mathrm{max}}$ for both dust populations were not changed during the fitting process.

Table\,\ref{tab:grid} summarizes the free parameters explored in our grid of models ($r_{\mathrm{sub}}$ is the radius at which the silicate dust grains reach the sublimation temperature of 1500\,K). For each source, we therefore computed a total of 25\,200 SEDs. To obtain the most probable values for all the parameters aforementioned, we computed probability distributions for all the free parameters. For each model, the goodness of the fit is assessed via a reduced $\chi^2_{\mathrm{r}}$ . These $\chi^2_{\mathrm{r}}$ values are then transformed into probabilities ($\propto$ exp[$- \chi^2_{\mathrm{r}}$ /2]), which are then projected onto each dimensions of the parameter space. This approach enables us to get a census on which parameters are well constrained (or not), as well as provide the most probable values. This methodology has already proven to be successfull for such studies (\citealp{Harvey2012}; \citealp{Spezzi2013}).

To be able to compare consistently models for different sources, the goodness of fit has to be estimated in a similar fashion for all 17 sources. Significant photometric variability is observed for several sources at near- and mid-IR wavelengths (e.g., {\it Spitzer}/IRAC \& MIPS versus WISE fluxes, see J11065906-7718535; Table\,\ref{tab:irac}), which is a main issue for SED modeling given that the photometric datapoints are non-simultaneous (e.g., \citealp{Joergens2012}). Several years have passed between the {\it Spitzer}, WISE, and 2MASS observations, and additionally {\it Spitzer}/IRAC, MIPS, and IRS data are usually not taken simultaneously. Photometric variability in the Class\,II phase can be caused for example by variable accretion (e.g., \citealp{Muzerolle2009}; \citealp{Joergens2012}) or occultation effects due to the disk (\citealp{Alencar2010}; \citealp{Looper2010}) and it affects mostly the near- and mid-IR fluxes and much less the far-IR. 

Out of the 17 sources, only two sources have not been observed with the IRS instrument (J10533978-7712338 and J11111083-7641574), while 14 sources have IRS spectra between 5 and 35\,$\mu$m and one source (J11062554-7633418) was only observed with the Short Low module (5.2-14.5\,$\mu$m). We therefore opted to use the IRS spectra to construct the $\chi^2_{\mathrm{r}}$, to minimize as much as possible the effect of time variability in the near- and mid-IR (i.e., IRAC, MIPS, and WISE measurements are not used when {\it Spitzer}/IRS spectra are available). For the 14 sources with a complete IRS spectra, we used five points at 5.5, 7.5, 9.8, 15, and 30\,$\mu$m (averaged over $\pm 0.1$\,$\mu$m). For J11062554-7633418, we used data from the IRS Short-Low module and used 4 points at 5.5, 7.5, 9.8, and 14\,$\mu$m, plus the WISE W4 point at 22\,$\mu$m. For the two sources without IRS data, we used the four WISE points and the IRAC point at 8\,$\mu$m. For all the sources, the near-IR excess is constrained by the 2MASS $K$-band photometric point. Finally, the uncertainties were set to 10\% of the observed fluxes for all wavelengths but the PACS wavelengths for which we used the uncertainties derived from the observations. The PACS upper limits were not included when computing the $\chi^2_{\mathrm{r}}$ values.

\section{Results\label{sec:res}}

\begin{figure}
\begin{center}
\includegraphics[width=\columnwidth]{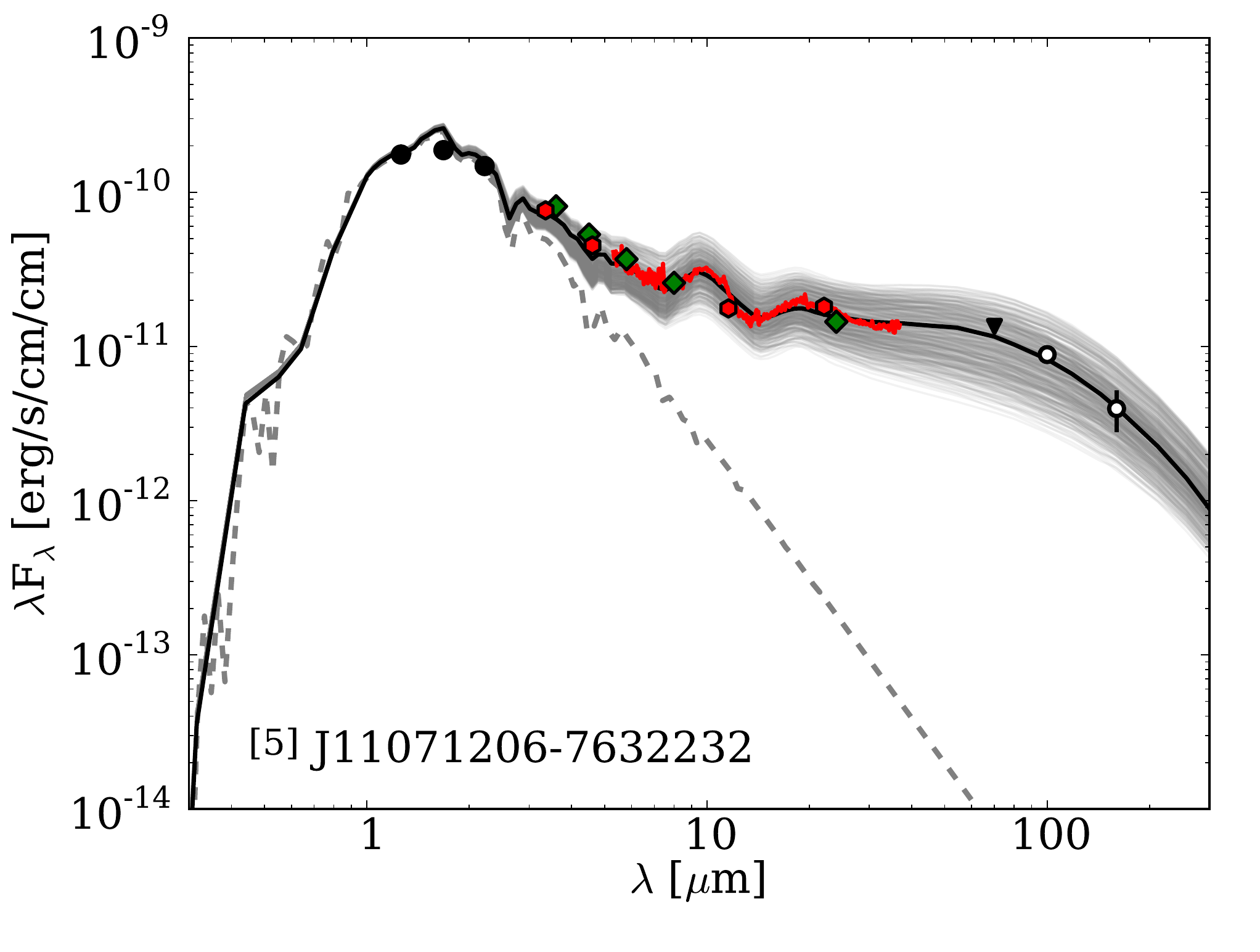}
\caption{SED of J11071206-7632232 (\#5). Grey dashed line represents the photosphere. Filled black circles are 2MASS measurements, green diamonds are {\it Spitzer}/IRAC and MIPS observations, red hexagons are the WISE measurements. {\it Spitzer}/IRS spectrum is shown in red. PACS observations are in open circles and upper limits are represented as filled black triangles. The most probable model is shown as a solid black line, and the grey lines correspond to models within the 68\% confidence interval (see text for details).\label{fig:SED_1}}
\end{center}
\end{figure}

\begin{table*}
\caption{Derived most probable values for the disk parameters of the 17 M-type stars. The second line indicates the range of validity for each parameter (see text for details).\label{tab:res}}
\begin{center}
\begin{tabular}{lcccccccc}
\hline \hline
Object Name & ID & $M_\mathrm{disk}$ & $r_\mathrm{in}$ & $H_0$ & $H_{0,\mathrm{small}}$ & $\gamma$ & $i$ & Spectral type \\
& & $[M_\odot]$ & $[r_\mathrm{sub}]$ & [AU] & [AU] & & [$\degr$] &  \\
\hline
J10533978-7712338 & 1 & 5$\times 10^{-3}$ & 3 & 10 & 5 & 1.30 & 68 & M2.75\\
 & & [5$\times 10^{-4}$, 5$\times 10^{-3}$] & [1,3] & [5,15] & [1,5] & [1.20,1.30] & [5,68] &  \\
J11044258-7741571 & 2 & $10^{-3}$ & 3 & 5 & 5 & 1.30 & 5 & M4\\
 & & [$10^{-3}$, 5$\times 10^{-3}$] & [2,3] & [5,5] & [2,5] & [1.25,1.30] & [5,32] &  \\
J11062554-7633418 & 3 & 5$\times 10^{-3}$ & 3 & 15 & 1 & 1.30 & 59 & M5.25\\
 & & [5$\times 10^{-3}$, 5$\times 10^{-3}$] & [1,3] & [15,25] & [0,2] & [1.25,1.30] & [32,59] &  \\
J11065906-7718535 & 4 & $10^{-3}$ & 3 & 10 & 2 & 1.15 & 50 & M4.25\\
 & & [5$\times 10^{-4}$, $10^{-3}$] & [1,3] & [10,10] & [1,5] & [1.10,1.15] & [14,59] &  \\
J11071206-7632232 & 5 & 5$\times 10^{-4}$ & 2 & 10 & 1 & 1.20 & 59 & M0.5\\
 & & [5$\times 10^{-4}$, $10^{-3}$] & [1,3] & [5,10] & [1,2] & [1.15,1.25] & [32,68] &  \\
J11074366-7739411 & 6 & 5$\times 10^{-4}$ & 1 & 10 & 0 & 1.10 & 23 & M0\\
 & & [5$\times 10^{-4}$, 5$\times 10^{-4}$] & [1,1] & [10,10] & [0,0] & [1.10,1.10] & [14,32] &  \\
J11085464-7702129 & 7 & $10^{-4}$ & 2 & 15 & 1 & 1.15 & 68 & M0.5\\
 & & [5$\times 10^{-5}$, $10^{-4}$] & [1,3] & [15,20] & [1,5] & [1.15,1.20] & [41,68] &  \\
J11091812-7630292 & 8 & 5$\times 10^{-5}$ & 1 & 15 & 1 & 1.15 & 50 & M1.25\\
 & & [5$\times 10^{-5}$, $10^{-4}$] & [1,2] & [15,20] & [1,2] & [1.10,1.15] & [32,59] &  \\
J11094742-7726290 & 9 & $10^{-3}$ & 2 & 5 & 5 & 1.25 & 50 & M3.25\\
 & & [5$\times 10^{-4}$, 5$\times 10^{-3}$] & [1,3] & [5,10] & [5,5] & [1.20,1.30] & [41,68] &  \\
J11095407-7629253 & 10 & 5$\times 10^{-4}$ & 3 & 15 & 1 & 1.20 & 68 & M2\\
 & & [5$\times 10^{-4}$, $10^{-3}$] & [2,3] & [10,15] & [1,2] & [1.15,1.20] & [50,68] &  \\
J11095873-7737088 & 11 & $10^{-3}$ & 1 & 5 & 5 & 1.00 & 23 & M1.25\\
 & & [5$\times 10^{-4}$, 5$\times 10^{-3}$] & [1,2] & [5,10] & [2,5] & [1.00,1.00] & [14,59] &  \\
J11100704-7629376 & 12 & $10^{-4}$ & 2 & 10 & 0 & 1.10 & 50 & M0\\
 & & [$10^{-4}$, $10^{-3}$] & [2,2] & [5,10] & [0,1] & [1.00,1.10] & [23,59] &  \\
J11104959-7717517 & 13 & 5$\times 10^{-4}$ & 3 & 15 & 5 & 1.15 & 5 & M2\\
 & & [5$\times 10^{-4}$, $10^{-3}$] & [2,3] & [10,15] & [2,5] & [1.15,1.15] & [5,23] &  \\
J11105333-7634319 & 14 & 5$\times 10^{-4}$ & 3 & 10 & 5 & 1.10 & 5 & M3.75\\
 & & [5$\times 10^{-4}$, $10^{-3}$] & [2,3] & [10,10] & [2,5] & [1.05,1.10] & [5,14] &  \\
J11105597-7645325 & 15 & $10^{-4}$ & 3 & 15 & 1 & 1.20 & 68 & M5.75\\
 & & [5$\times 10^{-5}$, $10^{-4}$] & [2,3] & [10,25] & [0,2] & [1.15,1.25] & [23,68] &  \\
J11111083-7641574 & 16 & $10^{-3}$ & 3 & 10 & 1 & 1.10 & 86 & M2.5\\
 & & [$10^{-3}$, $10^{-3}$] & [2,3] & [10,10] & [1,1] & [1.10,1.10] & [86,86] &  \\
J11120984-7634366 & 17 & $10^{-4}$ & 2 & 10 & 2 & 1.25 & 68 & M5\\
 & & [5$\times 10^{-5}$, 5$\times 10^{-4}$] & [1,3] & [5,15] & [1,5] & [1.20,1.30] & [23,68] &  \\
\hline
\end{tabular}
\end{center}
\end{table*}

\begin{figure*}
\begin{center}
\includegraphics[width=1.9\columnwidth]{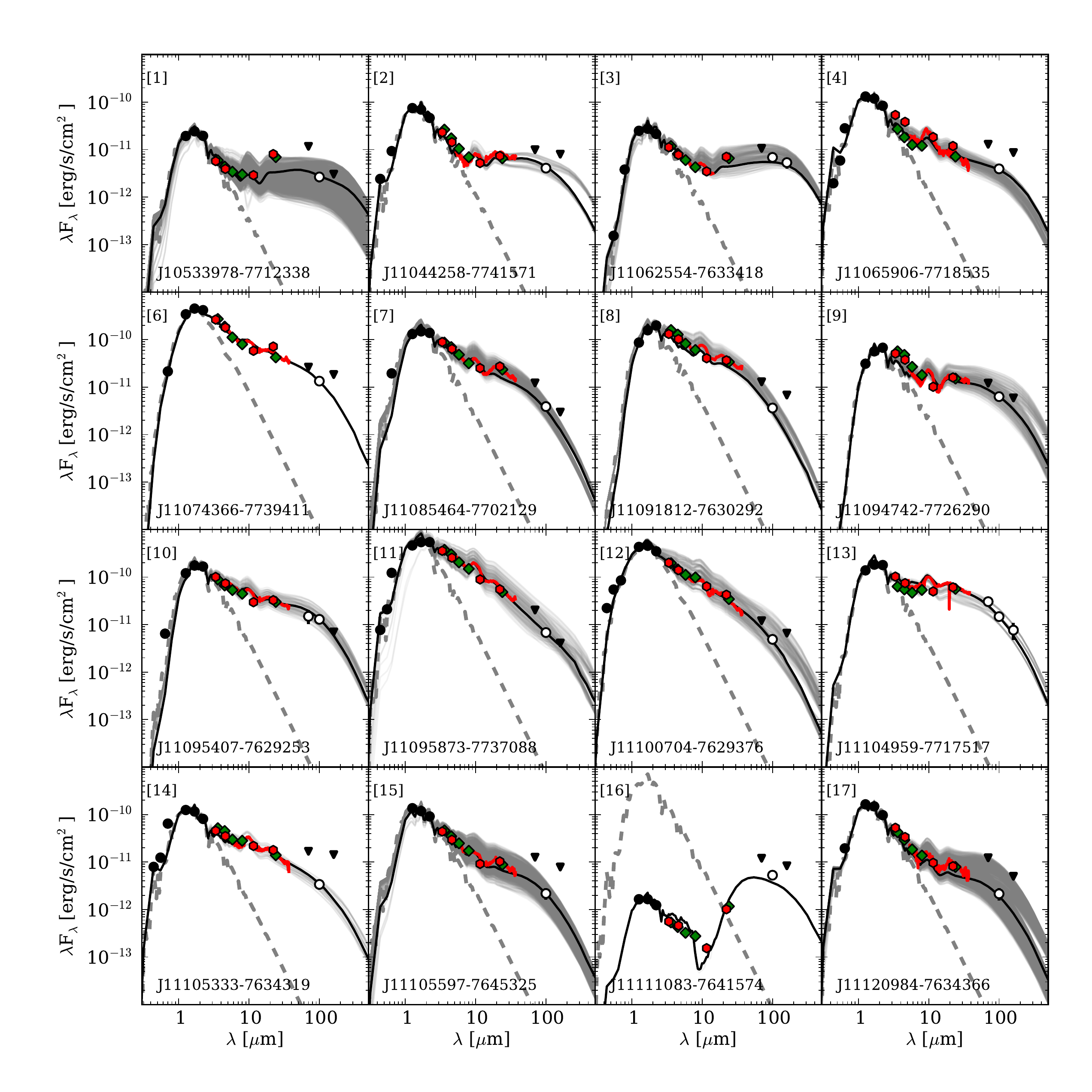}
\caption{Same as Figure\,\ref{fig:SED_1} for the 16 other M-type stars detected at PACS wavelengths.\label{fig:SED_2}}
\end{center}
\end{figure*}

We detected 17 sources in the far-IR belonging to the young star-forming region Cha-I and successfully model their SEDs. The most probable fits to the SEDs, as defined by models with disk parameters at the peak positions of their respective probability distributions, are displayed as solid black lines in Figures\,\ref{fig:SED_1} and \ref{fig:SED_2}. The most probable disk parameters are summarized in Table\,\ref{tab:res}. For a given source, the second line of the Table shows the range of validity for the considered parameter, obtained from the probability distribution. Given the distribution $P(\theta)$ for the parameter $\theta$ between $\theta_{\mathrm{min}}$ and $\theta _{\mathrm{max}}$, the range of validity $ [\theta_1,\theta_2] $ is found for $ \int_{\theta \mathrm{min}}^{\theta 1} P(\theta)\,\mathrm{d}\theta  = \int_{\theta 2}^{\theta \mathrm{max}} P(\theta)\,\mathrm{d}\theta = (1- \beta) /2$, where $\beta = 0.68$ (see e.g., \citealp{Pinte2008}). The interval $[\theta_1,\theta_2]$ corresponds to the 68\% confidence interval for each parameters. In Figures\,\ref{fig:SED_1} and \ref{fig:SED_2}, models that are within this interval are plotted in light grey.

\subsection{Edge-on disks and problematic sources\label{sec:edge-on}}

We could hardly find satisfying models for two sources: J10533978-7712338 and J11111083-7641574 (\#1 and 16, respectively). Source \#16 is most likely an edge-on disk, if we use the luminosity of 0.003\,$L_\odot$, as reported in the literature, the near-IR fluxes can be reproduced regardless of the disk inclination (as long as it is not too high), but the far-IR fluxes would be very much underestimated. The differences between the modeled SED and the far-IR observations could be reduced by setting the inclination to a higher value, and increasing the stellar luminosity to 0.8\,$L_\odot$ to match the near-IR data points. Even though our final model still slightly under-predicts the PACS measurement at 100\,$\mu$m, our finding compares well with the modeling presented in \cite{Robberto2012}, where the authors found $i \sim 87.1\degr$ for this source. 

The main issue in the modeling of source \#1 comes from the high fluxes at WISE W4 and MIPS 24\,$\mu$m wavelengths (59.2 and 55.4\,mJy, compared to our modeled fluxes of 26.4 and 28.4\,mJy at 22 and 24\,$\mu$m, respectively). First of all, detailled inspection of the individual WISE W4 images show the source is bright, clearly detected, and isolated from other nearby sources. The overall small reduced $\chi^2$ in the WISE catalog suggests it is a point source without any significant contamination. This is confirmed by the MIPS 24 point, which is close to the W4 point. However, no disk models could reproduce both the mid-IR datapoint and the very steep decrease toward the PACS measurement (detected at more than 5$\sigma$). The object is not located close to any bright region in the {\it Herschel} map and does not appear as extended either. \citet{Robberto2012} and \cite{Luhman2008} suggested the source is seen almost edge-on and is mainly seen in scattered light. In the following statistical analysis both problematic sources will be excluded from the sample.

\subsection{Detection statistics and probabilities\label{sec:det}}

\begin{table}
\caption{Detection statistics.\label{tab:det}}
\begin{center}
\begin{tabular}{lcc}
\hline \hline
Spectral type & Detections \# & Original sample \\
\hline
M0 & 4 & 4 \\
M1 & 2 & 8 \\
M2 & 4 & 5 \\
M3 & 2 & 8 \\
M4 & 2 & 8 \\
M5 & 3 & 18 \\
M6 & 0 & 4 \\
M7 & 0 & 1 \\
M8 & 0 & 3 \\
M9 & 0 & 3 \\
\hline
\end{tabular}
\end{center}
\end{table}

The original stellar sample from \citet{SzHucs2010} contained 62 M-type stars, and 17 sources out of these 62 were successfully detected at least at one PACS wavelengths. Table\,\ref{tab:det} summarizes the number of objects detected and the total number of sources in the stellar sample for different spectral types from M0 to M9. First of all, while it is expected to not detect very low-mass objects, it appears that several M1 stars of the original list are not detected by the PACS instrument (25\% detection rate). Close inspection of the 100\,$\mu$m PACS map reveals that several of these M1 stars are close to bright, extended structures, rendering any detection impossible.

\begin{figure}
\begin{center}
\includegraphics[width=1\columnwidth]{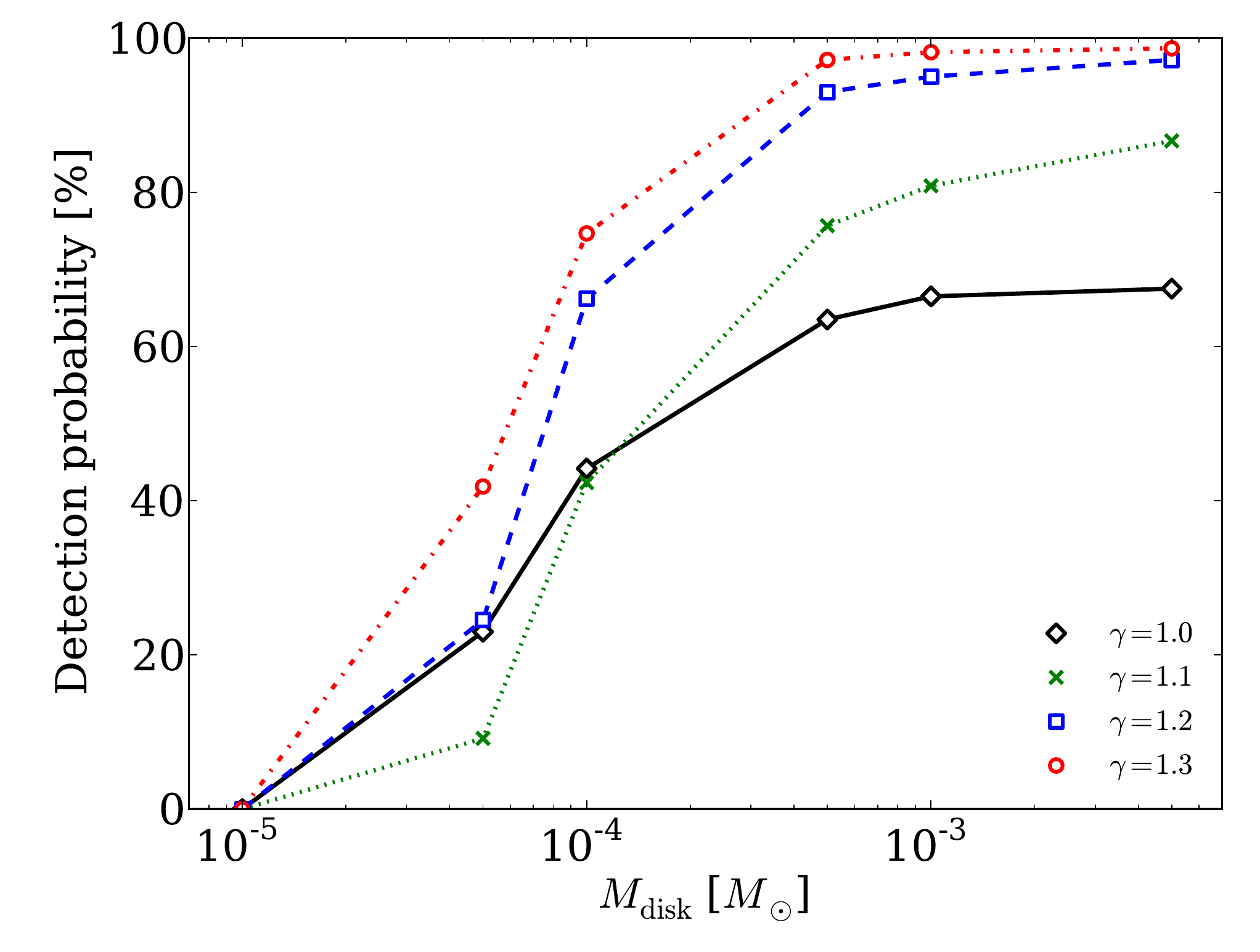}
\caption{Detection probabilities at $\lambda = 100\,\mu$m, based on the grid of models for the M5 star J11120984-7634366 (\#17), as a function of the disk mass for different flaring indices (see text for details).\label{fig:det}}
\end{center}
\end{figure}

Based on the numerous SEDs modeled within the grids, we can better investigate the probabilities of detecting a disk for a certain set of parameters. Because the 100\,$\mu$m map is the most sensitive of all three PACS bands, we considered a 3\,$\sigma$ detection threshold of 60\,mJy to be representative of the observations (see Table\,\ref{tab:pacs}). We then estimated the 100\,$\mu$m fluxes for all the models of a single source (e.g., source \#17 for a M5 star) and computed the fraction of models that would be detected as a function of certain disk parameters. Figure\,\ref{fig:det} shows the detection probabilities as a function of the disk mass, where different values for the flaring indices $\gamma$ are color-coded, for the M5 star J11120984-7634366. The fact that for small disk masses the detection probabilities are larger for $\gamma = 1.0$ compared to $\gamma = 1.1$ is not a numerical issue in the radiative transfer calculations. We tested this against the number of photons in the model, and the resolution of the radial grid. The fact that the detection probabilities become smaller for $\gamma = 1.0$ at higher masses tends to indicate that different regions in the disks are probed at 100\,$\mu$m for small disk masses (optical depth effect). Besides this, one can immedialety see that the PACS observations at 100\,$\mu$m are not sensitive enough to detect disk masses of about $10^{-5}$\,$M_{\odot}$ around this star. These simple considerations enable us to better understand the detection rates presented in Table\,\ref{tab:det}. All the M0 stars of our stellar sample were detected because no matter how massive, or how flared, the disk is, the chances of detecting it are overall large, while the probabilities of detecting relatively low-mass, reasonably flared disks (e.g., $M_{\mathrm{disk}} = 10^{-4}$\,$M_{\odot}$, $\gamma \sim 1.1$) around M5 stars rapidly drop below 50\%. According to our modeling results, all the three M5 stars detected in our sample have high flaring indices ($\gamma = 1.3$ for J11062554-7633418 and J11120984-7634366, and $1.2$ for J11105597-7645325), which should {\it not} be interpreted as an evidence for larger $\gamma$ values for disks around M5 stars, in addition to the overall challenge when dealing with low-number statistics.

The stellar sample considered in this study originally aimed at selecting disk-bearing sources (\citealp{SzHucs2010}), based on their near-IR excesses. However, 15 M5-type stars that are known to harbor a disk were not detected in the {\it Herschel} observations. \citet[][and references therein]{Manoj2011} reported that five M5 stars in Cha\,I are in multiple systems, some of them having small separations. For instance 0.17$''$ separates the primary and secondary of the Cha\,Ha2 system (\citealp{Ahmic2007}; \citealp{Lafreni`ere2008a}) corresponding to about 28\,AU at 165\,pc. Such small separation can have a strong impact on the disk (see e.g., \citealp{Bitner2010}), and may explain why the disk is not detected at PACS wavelengths. However, out the of five M5 stars with known multiplicity, two of them are detected with the {\it Herschel} data even at small separation: objects \#6 (28.9$''$ separation, \citealp{Kraus2007}) and \#15 (0.13$''$ separation \citealp{Ahmic2007}; \citealp{Lafreni`ere2008a}). It does not seem (based on five sources) that binarity can solely explain the small detection rate of M5 stars in the {\it Herschel} map of Cha-I. Even though the detection probabilities shown in Figure\,\ref{fig:det} are derived for a single source, with its own luminosity, this would suggest that the disks around these 15 undetected M5 stars are most likely low-mass ($M_{\mathrm{disk}} \lesssim 5 \times 10^{-5}$\,$M_\odot$ for a flaring index of about $\gamma \sim 1.1$), or have low flaring indices ($\gamma \leq 1.1$) for slightly higher disk masses, the two parameters being most likely degenerate.

\subsection{Disk masses and disk evolution\label{sec:masses}}

As discussed in Sect.\,\ref{sec:det}, we most likely detected the most massive, or more flared disks around M3-M5 stars. This detection bias renders the question of disk masses around M-type stars challenging at the sensitivity level of the {\it Herschel} maps. Nonetheless, our study complements the current effort towards a better characterization of disks around low-mass stars (e.g., \citealp{Harvey2012}; \citealp{Joergens2012}). Our sample contains slightly more massive objects and fills in the gap for several M0--M3 type stars. We find that disks around such objects have masses in the range $10^{-4}$--$10^{-3}$\,$M_{\odot}$. Based on the detection probabilities as a function of PACS sensitivity at 100\,$\mu$m, we already concluded that disks around M5 stars (or later spectral types) should either {\it (i)} have masses of the order of $10^{-5}$\,$M_{\odot}$, or below, a result consistent with the study from \citet{Harvey2012}, or {\it (ii)} appear flatter than early M-type stars (low $\gamma$ values). Additionally, we searched for a correlation between the estimated disk masses and the 100\,$\mu$m fluxes. To estimate the significance (or lack) of the correlation between the two quantities, we used the Kendall $\tau$ correlation coefficient (between $1$ and $-1$, for a perfect correlation and anti-correlation, respectively), and its associated probability $P$ (the smaller, the more robust). We obtained a Kendall $\tau$ correlation coefficient of 0.23, with a probability $P = 0.22$. The lack of a correlation implies that the measured far-IR fluxes are only a reliable measure by themselves for disk masses in the regime $M_{\mathrm{disk}} \lesssim 10^{-5}$\,$M_\odot$ where a larger fraction of the total disk mass lies in regions of smaller optical depths in the far-IR (\citealp{Harvey2012}). This result implies that, as opposed to mm observations, fluxes in the far-IR cannot be {\it directly} used to measure disk masses around M0-M5 type stars. A careful SED modeling effort is the next best approach to estimate disk masses in the higher mass regime.

\begin{figure}
\begin{center}
\includegraphics[width=\columnwidth]{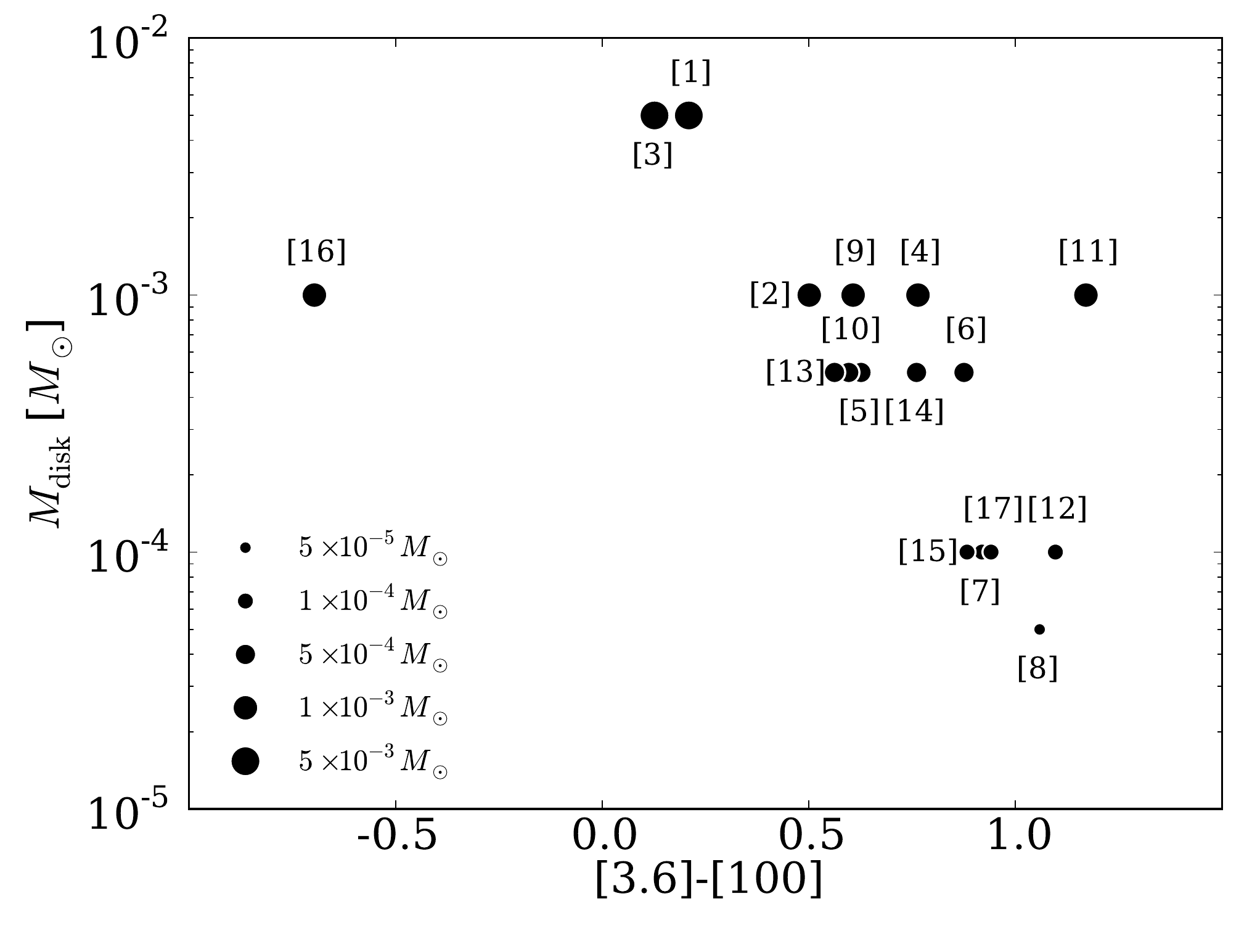}
\caption{Most probable disk masses $M_{\mathrm{disk}}$ as a function of near- to far-IR color ($[3.6]-[100]$).\label{fig:color_mass}}
\end{center}
\end{figure}

In an attempt to link the most probable disk masses to the evolution of the circumstellar disk, we show in Figure\,\ref{fig:color_mass} the disk masses inferred from the modeling as a function of the near- to far-IR color ($[3.6]-[100]$), quantity defined as 
\begin{eqnarray}
[a] - [b] = - \frac{\mathrm{log}(\lambda_{b} F_{b}) - \mathrm{log}(\lambda_{a} F_{a})}{\mathrm{log}(\lambda_{b}) - \mathrm{log}(\lambda_{a})},
\end{eqnarray}
where $F_{a,b}$ is the flux density in units of erg.s$^{-1}$.cm$^{-2}$.$\mu$m$^{-1}$ at wavelength $\lambda_{a,b}$ in units of $\mu$m. As the near- to far-IR colors become bluer (increasing $[3.6]-[100]$), the modeled disk masses become smaller, a result in line with (sub-)mm surveys of young stellar objects (e.g., \citealp{Andrews2007}). Including all our sources, we find a Kendall $\tau$ correlation coefficient of $-0.58$ with an associated probability $P = 1.1 \times 10^{-3}$. Consequently, even if the most probable disk masses do not correlate well with the measured {\it Herschel} fluxes, they do anti-correlate relatively well with the near- to far-IR shape of the SEDs, a quantity that can be related to the far-IR opacity of the disks (disk evolution or substantial growth of dust particles, \citealp{Beckwith1990}; \citealp{Andrews2007}).

\subsection{Disk flaring}

\begin{figure}
\begin{center}
\includegraphics[width=\columnwidth]{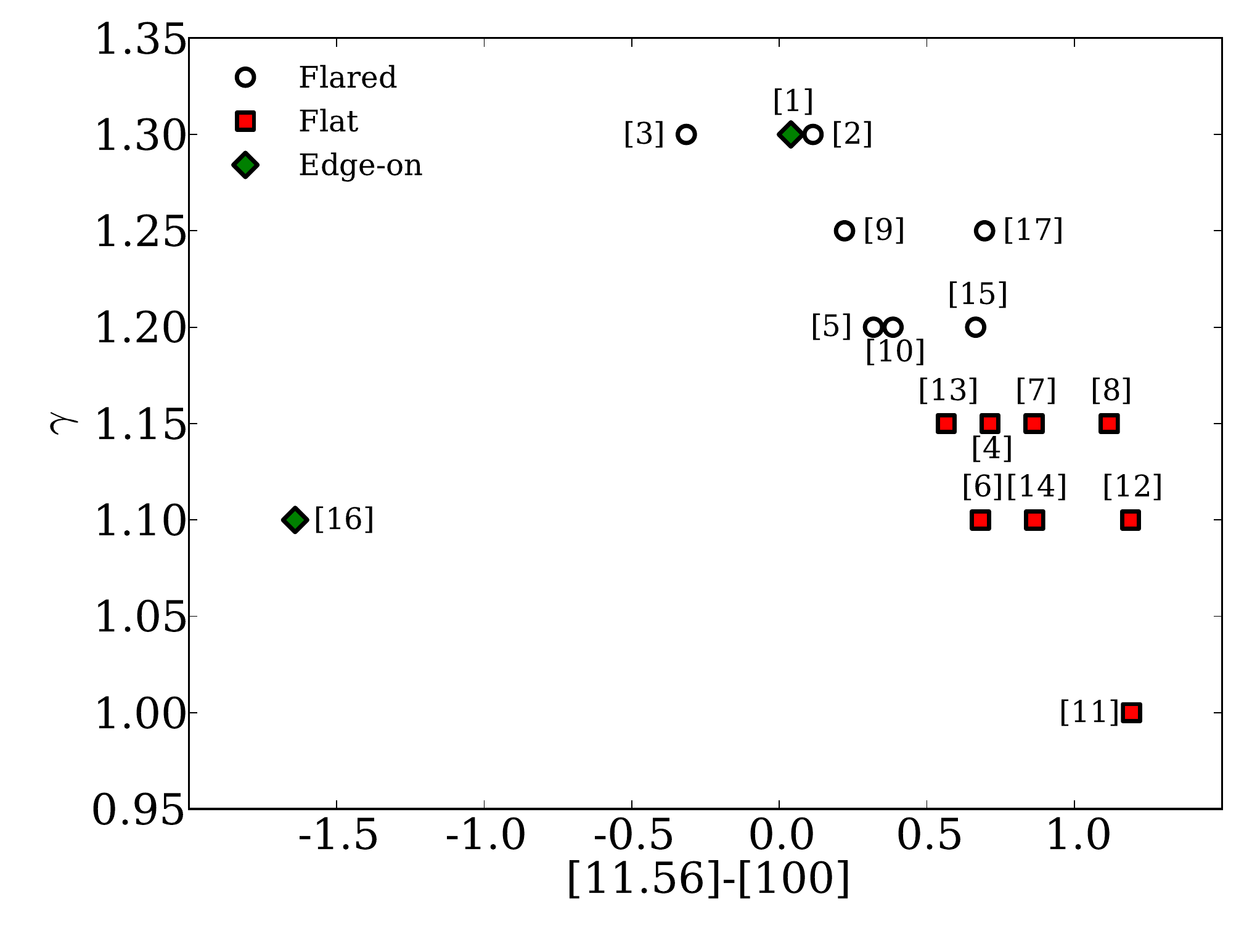}
\caption{Most probable flaring indices $\gamma$ as a function of mid- to far-IR color ($[11.56]-[100]$).\label{fig:color_gamma}}
\end{center}
\end{figure}

Our modeling results point toward a relatively broad distribution of flaring indices ($1 \leq \gamma \leq 1.3$). We can therefore search for a first order approximation between the SEDs and the flaring indices. Figure\,\ref{fig:color_gamma} shows the mid- to far-IR colors (between the WISE band W3 and the 100\,$\mu$m flux) as a function of the most probable flaring indices $\gamma$. In Figure\,\ref{fig:color_gamma}, sources with $\gamma \geq 1.2$ are shown with open black circles, and $\gamma \leq 1.15$ with red squares. The two sources that are most likely edge-on (\#1 and \#16) are shown as green diamonds. Excluding the two edge-on sources from the sample (\#1 and \#16), we find a correlation coefficient $\tau = 0.73$ with the associated probability $P=1.6 \times 10^{-4}$, suggesting the two quantities correlate well. We find that a threshold value $[11.56]-[100] \sim 0.5$ appears to separate relatively well flat and flared disks.

\subsection{A tentative characterisation of dust vertical distribution}

The {\it Spitzer}/IRS observations are probing the optically thin upper layers of the circumstellar disks. Even though we did not aim at fitting the emission features perfectly, our parametrization of the dust stratification (via the parameter $H_{0,\,\mathrm{small}}$) has proven to be a good approach to characterize dust settling and reproduce the overall shape of the 10\,$\mu$m emission feature (especially their peak fluxes). For all the sources but two (sources \#6 and 12), a population of small dust grains in the uppermost layers is required to provide a satisfaying match to the IRS data, even though the confidence intervals for $H_{0,\,\mathrm{small}}$ are fairly large (except for source \#9 which has the strongest 10\,$\mu$m feature in our sample). The IRS spectra of sources \#6 and 12 do show a 10\,$\mu$m emission feature, but the features appear slightly weaker and flatter than for some of the other sources (e.g., sources \#9, 11 or 13).

To investigate deeper this possible relation, we computed the ratio between optically thin and thick emission at 10\,$\mu$m, from the IRS spectra. As a first order approximation, the emission feature at 10\,$\mu$m indeed arises from optically thin regions of the disk, while the continuum emission originates from optically thick regions. We therefore defined a linear, local continuum ($F_{\nu,\mathrm{cont}}$) underlying the observed 10\,$\mu$m emission feature ($F_{\nu,\mathrm{obs}}$) between $\lambda_1 = 7.5$ and $\lambda_2 = 13.5$\,$\mu$m. The observed ratio between optically thin and thick emission can then be approximated by the quantity $\int_{\lambda_1}^{\lambda_2} (F_{\nu,\mathrm{obs}} - F_{\nu,\mathrm{cont}}) \, \mathrm{d}\lambda/ \int_{\lambda_1}^{\lambda_2} F_{\nu,\mathrm{cont}}\,\mathrm{d}\lambda$. We find that this quantity may correlate tentatively with the $H_{0,\mathrm{small}}$ parameter for the most probable fits to the SED (Kendall $\tau = 0.39$, with the associated probability $P=0.05$), but there is no sign of correlations with other parameters such as $M_{\mathrm{disk}}$, $H_0$ or $\gamma$. By adding more small grains in the optically thin regions of the disk, the parametrization of the dust stratification seems to be a good solution to reproduce strong 10\,$\mu$m emission features as the one of object \#9. For the other sources of the sample, the situation is more ambiguous (hence the small correlation Kendall $\tau$ coefficient). Computing the value $S_{\mathrm{peak}}$ from the IRS spectra, as in \citet{Kessler-Silacci2006}, to quantify the amount of dust processing, we find a median value of 1.7 with a standard deviation of 0.7, indicating ``boxy'' emission features attributed to $\mu$m-sized grains or shallow grain size distributions (\citealp{Olofsson2010}). These values match the vast majority of previous {\it Spitzer}/IRS observations of young circumstellar disks (e.g., \citealp{Bouwman2008}; \citealp{Furlan2009}; \citealp{Olofsson2009}; \citealp{Manoj2011}; \citealp{Oliveira2011}). However, while we aimed for reproducing the peak flux of the 10\,$\mu$m feature, we did not aim for matching their shape (grain size distribution and crystallinity), a question beyond the scope of this paper. 

Nonetheless, characterizing dust settling based on SED modeling is a challenging problem. \cite{Dullemond2008} have shown that dust sedimentation can have an effect on the 10\,$\mu$m emission feature when the grain size distribution is bimodal (two different grain sizes), but this effect is severely damped when considering a continuous grain size distribution. Our approach is closer to the bimodal distribution as we do not have a full description of dust settling. Consequently, the question of dust stratification in these disks will have to be further investigated, for instance by observing a flat sub-mm slope confirming the presence of mm-sized dust grains in the disk midplane.

\subsection{Sub-mm LABOCA observations}

\begin{figure}
\begin{center}
\includegraphics[width=\columnwidth]{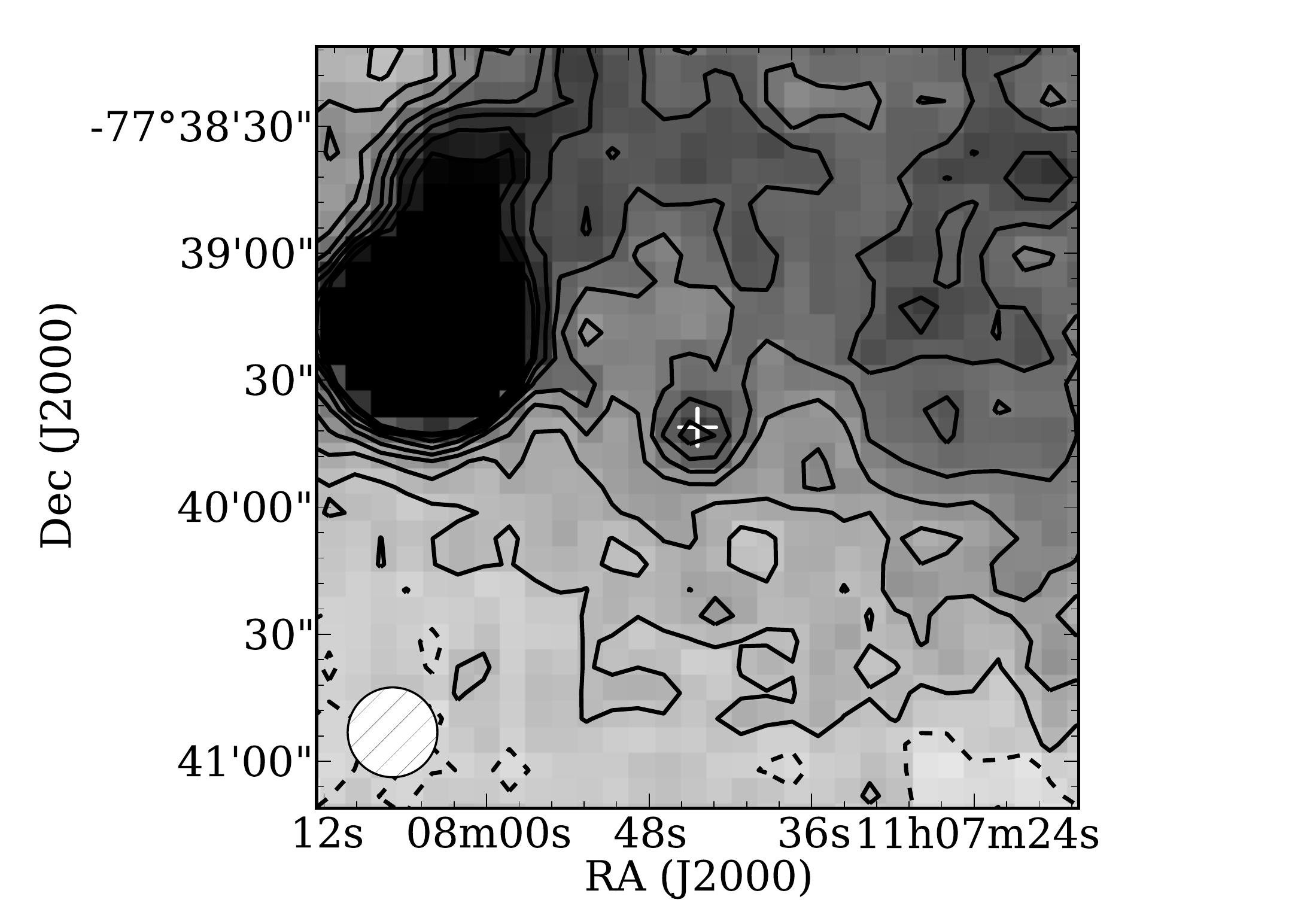}
\caption{LABOCA observations of source \#\,6 (marked with a white ``+'' symbol), which suggest the measured flux is contaminated by the nearby source. The beam size is displayed in the lower left corner.\label{fig:laboca6}}
\end{center}
\end{figure}

The Cha-I star forming region has been observed in the sub-mm with the bolometer array LABOCA at the APEX telescope (\citealp{Belloche2011}). Out of the 17 sources detected with {\it Herschel}/PACS, three sources were also detected at 870\,$\mu$m (sources \#\,6, 9, and 16). The fluxes reported by \cite{Belloche2011} are 94, 132, and 72\,mJy for these three sources, respectively. Inspecting the LABOCA map, we find the flux for source \#\,6 (J11074366-7739411, see Fig.\,\ref{fig:laboca6}) may suffer from contamination by a nearby source. While the close-by source is also detected in the 100\,$\mu$m PACS map, it is much more compact and does not contaminate the far-IR measurement. Therefore, we estimated the peak flux in the 870\,$\mu$m map by fitting a Gaussian profile with a full width at half maximum of 21.2$''$ (equal to the beam size) on top of the background emission, using the IDL package \texttt{Starfinder} (\citealp{Diolaiti2000}). We obtain a peak flux of about $\sim$\,58\,mJy (instead of 94\,mJy). Our most probable models severely under-predict the observed sub-mm fluxes, with predicted fluxes of 10, 14 and 10\,mJy for sources \#\,6, 9, and 16, respectively.

The first possible explanation for these discrepancies is that we under-estimate the disk masses with our models. \citet{Belloche2011} reported masses of $4.4 \times 10^{-3}$, $1 \times 10^{-2}$, and $5.5 \times 10^{-3}$\,$M_\odot$ for sources \#\,6, 9 and 16, respectively (assuming $F_{\nu} = 58$\,mJy at $\lambda = 870$\,$\mu$m for source \#\,6), a factor 5 to 10 above our results. In Section\,\ref{sec:masses}, we already discussed that the measured PACS fluxes do not correlate with the inferred disk masses, indicating that the emission is not fully optically thin at 100\,$\mu$m. Consequently, the {\it Herschel} observations are not direct tracers of the amount of dust in the disks we modeled (hence the effort of modeling the SED from optical to far-IR wavelengths, see also Sect.\,\ref{sec:discuss} for further discussion). However, the purpose of this study is to compare relative disk parameters for stars of different spectral types, modeled in a consistent way. Furthermore, the fluxes differences in the sub-mm can also be the consequences of several assumptions that have to be made. First, as demonstrated by \cite{Draine2006}, the opacity in the sub-mm wavelength regime highly depends on the maximum grain sizes, the grain size distribution and the dust composition. The dust content is described by the optical constants of astronomical silicates, while carbonaceous grains, icy grains could well be present in the outermost regions of the disk. This would in turn modify the sub-mm slope of our diks models. We choose $s_{\mathrm{max}} = 1$\,mm, which is large enough at PACS wavelengths but may not be the best choice to model observations at 870\,$\mu$m (following the criterion of $s_{\mathrm{max}} \geq 3 \lambda$ of \citealp{Draine2006}). Increasing the maximum grain size to a few mm may slightly increase the sub-mm fluxes of our models. Additionally, far-IR measurements may not be sensitive to cold dust in the outermost regions, suggesting that regions of the disk may remain unseen by {\it Herschel}. Finally, optical constants are usually measured at room temperature, while the dust grains in the outermost regions of circumstellar disks can be as cold as a few tens of Kelvin. \cite{Coupeaud2011} have demonstrated that the temperatures of the grains during laboratory experiments have drastic effects on their opacities, especially in the sub-mm. Investigating these effects for three sources in our sample with sub-mm data is out of the scope of this study.

\section{Discussion\label{sec:discuss}}

\begin{figure*}
\begin{center}
\includegraphics[width=\columnwidth]{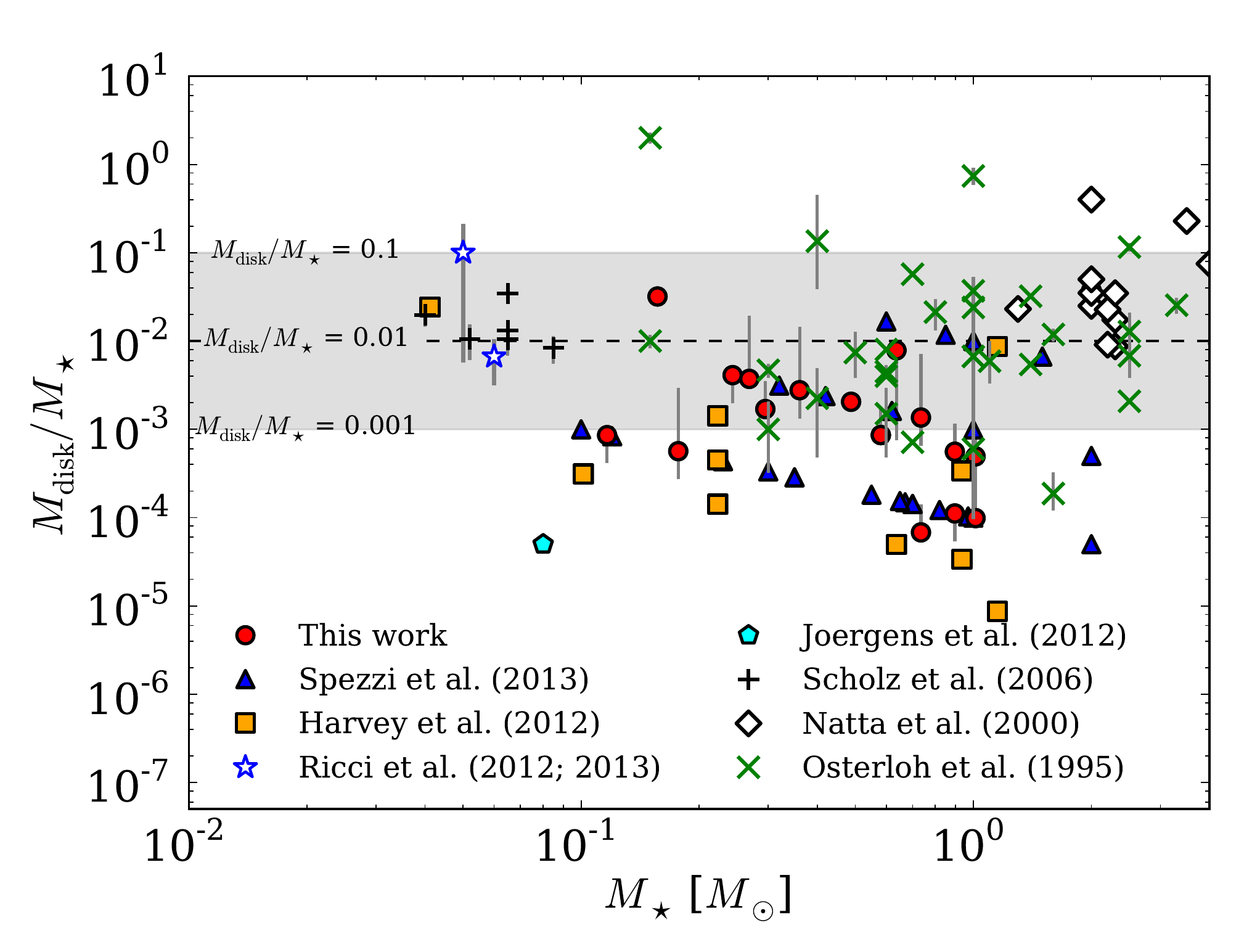}
\includegraphics[width=\columnwidth]{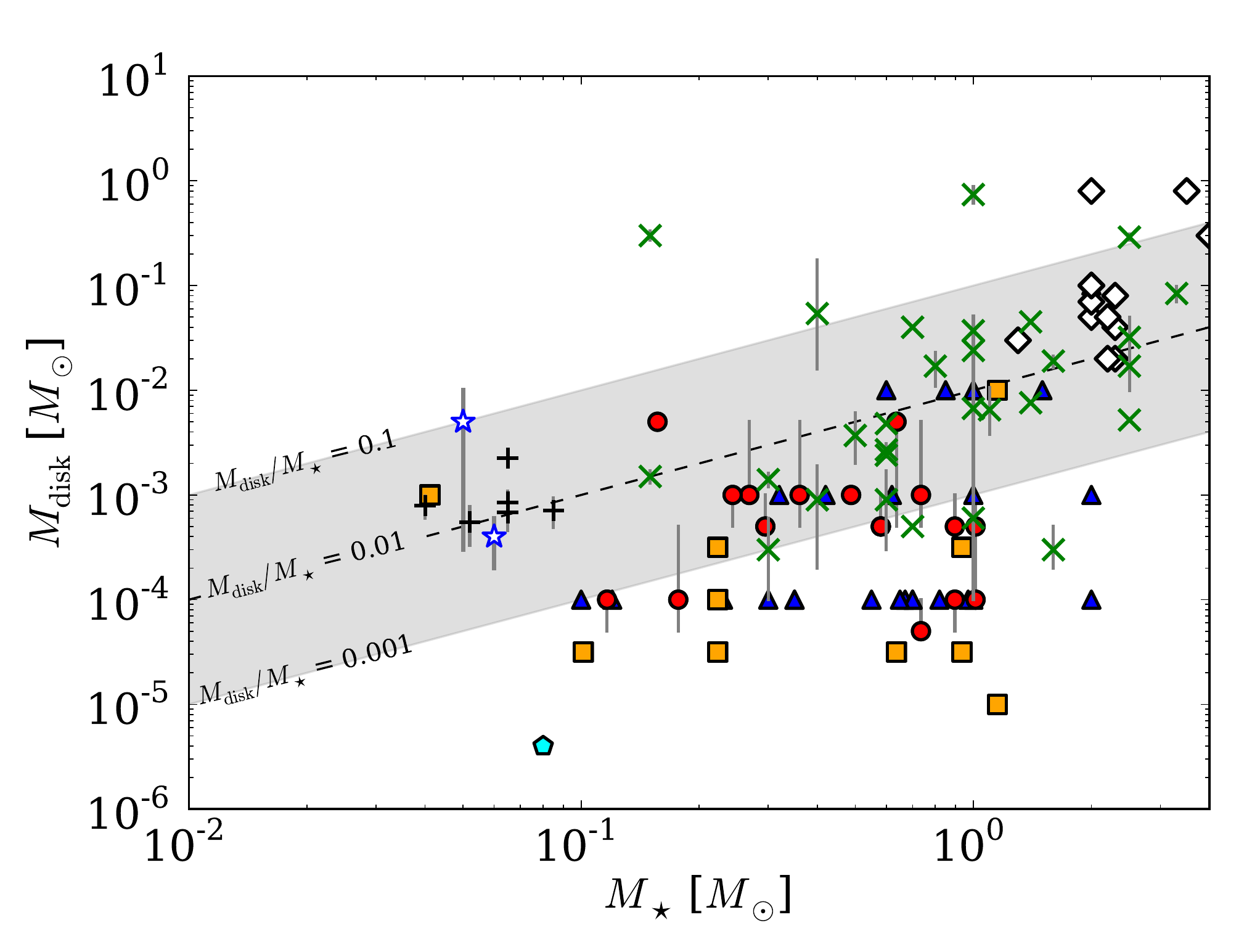}
\caption{Disk mass to stellar mass ratio ({\it left panel}) and disk mass ({\it right panel}) as a function of stellar mass. On each panel, the shaded area represents $\mathrm{log} (M_{\mathrm{disk}}/M_{\star}) = -2 \pm 1$. Vertical grey lines correspond to uncertainties, when available. Data represented with green ``$\times$'' symbols from \cite{Osterloh1995}, open black diamonds from \cite[][and references therein]{Natta2000}, dark ``+'' symbols are from \cite{Scholz2006}, cyan hexagon from \citet{Joergens2012}, open blue stars from \citet{Ricci2012,Ricci2013}, orange squares from \citet[][Cha-I sample only]{Harvey2012}, blue triangles from \citet[][lower limits]{Spezzi2013}, and open red circles from this work. The color-coding in right panel is the same as in left panel. Filled symbols correspond to disk masses inferred from modeling of far-IR observations (this work, \citealp{Joergens2012}, \citealp{Harvey2012}, and \citealp{Spezzi2013}).
\label{mdisk_mstar}}
\end{center}
\end{figure*}

Several studies have speculated that some disk properties are different for disks around Sun-like stars and very-low mass objects (e.g., \citealp{Pascucci2009}). The two categories being usually defined for spectral types between G5--K1 to M4.5 for the first group and M5 to M9 for the latter group. For instance, based on {\it Spitzer}/IRACs color distributions of a statistically large sample, \cite{SzHucs2010} concluded that disks may appear to be flatter around cooler stars, a consequence of dust settling towards the disk midplane. While the low-number statistics and the biases inherent to the sensitivity of the {\it Herschel}/PACS observations prevent us from drawing firm conclusions our findings can still be put into a broader picture by comparing them to other studies.

Mainly, the question of the available mass in disks around low-mass stars has motivated several observational campaigns in the past years, to better assess the efficiency of planet formation in such systems. The underlying motivation being to assess if the disk mass is a function of the central stellar mass. Therefore, to compare the disk masses inferred in our analysis as a function of $M_\star$, we gathered results from the literature, for a broader range of stellar masses. First, for our 17 M-type stars, we estimated the masses of the host stars by interpolating isochrones from \citet[][assuming an age of 2\,Myr]{Baraffe2003} over their effective temperatures $T_{\mathrm{eff}}$ (see Table\,\ref{tab:stellar}). In Figure\,\ref{mdisk_mstar}, the ratio $M_{\mathrm{disk}}/M_\star$ as a function of the stellar mass is shown in the left panel and the relation $M_{\mathrm{disk}}$ as a function of $M_\star$ is displayed in the right panel. We show results from our analysis and from the literature (\citealp{Osterloh1995}; \citealp{Natta2000}; \citealp{Scholz2006}; \citealp{Joergens2012}; \citealp{Ricci2012,Ricci2013}; \citealp{Harvey2012}; \citealp{Spezzi2013}). The stellar masses for the Cha-I sample from \citet{Harvey2012} were determined from the stellar effective temperatures, in a similar way as for our sample. One should keep in mind that such a representation is affected by detection biases, especially for very low-mass stars. But overall, no clear trends are visible, as there is a significant vertical dispersion at all stellar masses, a conclusion in line with the results of \citet{Scholz2006} and \citet{Williams2011}. In both panels of Figure\,\ref{mdisk_mstar} the shaded area correspond to the trend $\mathrm{log} (M_{\mathrm{disk}}/M_{\star}) = -2 \pm 1$ observed in (sub-)mm surveys (e.g., \citealp{Williams2011}). It appears that several disk masses inferred from SED modeling of far-IR observations (filled symbols) fall below the trend determined from (sub-)\,mm observations. This may indicate that optical depth effects are still important in the far-IR and that the modeled disk masses may be lower limits estimates, keeping in mind that far-IR observations may not be sensitive to cold dust in the outermost regions. One can note from Figure\,\ref{mdisk_mstar} that the dispersion in disk masses is smaller for low mass stars than for sources with solar masses, which may be related to lower far-IR optical depths for disks around low mass stars. Overall, the {\it absolute} values of $M_{\mathrm{disk}}$ seem to be smaller for disks around (very) low-mass stars, indicating that massive planets (a few $M_{\mathrm{Jup}}$) can hardly form in these disks (e.g., \citealp{Mordasini2012}). However, the {\it relative} $M_{\mathrm{disk}}/M_\star$ ratio appears to remain constant over a large range of stellar masses, a valuable input information for planet formation theories. Future ALMA observations will provide more accurate mass determination by probing optically thin emission, with a better sensitivity, and will enable us to investigate deeper the stellar mass dependance.

\section{Conclusion\label{sec:conclusion}}

We studied and modeled the SEDs of 17 disk-bearing, low-mass M-type stars (spectral types between M0 and M5.75) detected at {\it Herschel}/PACS wavelengths in Cha-I (out of 62 M-type stars). The increased SED coverage at far-IR wavelength, combined with a careful modeling approach, enabled us to discuss further the disk masses, flaring indices and dust settling in regions of the disks responsible for emission up to far-IR wavelengths.

First, we discussed the detection rates of known disk-bearing sources (based on near-IR excess) and concluded the {\it Herschel} sensitivity for these observations not to be sufficient to detect disk masses below $\sim 10^{-5}$\,$M_\odot$, or flaring indices below $\gamma \leq 1.1$, for disks around M5 stars (or later spectral types). For sources detected in the PACS map, we find a distribution of disk masses between $10^{-4}$--$10^{-3}$\,$M_\odot$, which compares well with other studies of similar sources in the far-IR (e.g., \citealp{Harvey2012}; \citealp{Spezzi2013}). We find that the most probable disk masses anti-correlate with the near- to far-IR color of the SED. However, the far-IR fluxes do not correlate well with the derived disk masses, suggesting the contribution of regions with high optical depth, which are often non negligeable even at these wavelengths. Consequently, for several sources in our sample, we find that the most probable masses lie below the trend $\mathrm{log}(M_{\mathrm{disk}} / M_\star ) = -2 \pm1$ observed in (sub)-mm surveys of young disks. This suggests the disk masses found via modeling of far-IR observations can most likely be considered as lower limits estimates. Future ALMA observations will provide more reliable disk masses estimates (e.g., \citealp{Ricci2012,Ricci2013}). Finally, we suggest that dust settling may already take place in these young disks ($\sim 2$\,Myr), via the analysis of the 10\,$\mu$m emission feature, and especially its strength with respect to the underlying continuum. The parametrized stratification enables us to constrain the presence of small dust grains in the uppermost optically thin layers of the disks, while the observations still remain un-sensitive to the grain sizes in the disk midplane. Sub-mm observations are expected to provide further constraints on the dust grain sizes in regions closer to the midplane (e.g., \citealp{Boehler2013}).

\begin{acknowledgements}
The authors thank the anonymous referee for constructive comments that improved the paper and the discussion about the disk masses. L. Sz. acknowledges support from the Deutsche Forschungsgemeinschaft via SFB project 881 ``The Milky Way System'' (sub-project B3). The radiative transfer models were calculated on the bwGRiD (http://www.bw-grid.de), member of the German D-Grid initiative, funded by the Ministry for Education and Research (Bundesministerium f\"ur Bildung und Forschung) and the Ministry for Science, Research and Arts Baden-W\"urttemberg (Ministerium f\"ur Wissenschaft, Forschung und Kunst Baden-W\"urttemberg) and KOLOB cluster at the Heidelberg University. The authors thank Paul M. Harvey and Christophe Pinte for providing the disk masses of their Cha-I sample, and Amelia Bayo for fruitful discussions. This research has made use of the SIMBAD database, operated at CDS, Strasbourg, France.
\end{acknowledgements}

\bibliography{biblio}

\begin{thebibliography}{67}
\expandafter\ifx\csname natexlab\endcsname\relax\def\natexlab#1{#1}\fi

\bibitem[{{Ahmic} {et~al.}(2007){Ahmic}, {Jayawardhana}, {Brandeker}, {Scholz},
  {van Kerkwijk}, {Delgado-Donate}, \& {Froebrich}}]{Ahmic2007}
{Ahmic}, M., {Jayawardhana}, R., {Brandeker}, A., {et~al.} 2007, \apj, 671,
  2074

\bibitem[{{Alencar} {et~al.}(2010){Alencar}, {Teixeira}, {Guimar{\~a}es},
  {McGinnis}, {Gameiro}, {Bouvier}, {Aigrain}, {Flaccomio}, \&
  {Favata}}]{Alencar2010}
{Alencar}, S.~H.~P., {Teixeira}, P.~S., {Guimar{\~a}es}, M.~M., {et~al.} 2010,
  \aap, 519, A88

\bibitem[{{Alibert} {et~al.}(2011){Alibert}, {Mordasini}, \&
  {Benz}}]{Alibert2011}
{Alibert}, Y., {Mordasini}, C., \& {Benz}, W. 2011, \aap, 526, A63

\bibitem[{{Andrews} \& {Williams}(2005)}]{Andrews2005}
{Andrews}, S.~M. \& {Williams}, J.~P. 2005, \apj, 631, 1134

\bibitem[{{Andrews} \& {Williams}(2007)}]{Andrews2007}
{Andrews}, S.~M. \& {Williams}, J.~P. 2007, \apj, 671, 1800

\bibitem[{{Andrews} {et~al.}(2009){Andrews}, {Wilner}, {Hughes}, {Qi}, \&
  {Dullemond}}]{Andrews2009}
{Andrews}, S.~M., {Wilner}, D.~J., {Hughes}, A.~M., {Qi}, C., \& {Dullemond},
  C.~P. 2009, \apj, 700, 1502

\bibitem[{{Baraffe} {et~al.}(2003){Baraffe}, {Chabrier}, {Barman}, {Allard}, \&
  {Hauschildt}}]{Baraffe2003}
{Baraffe}, I., {Chabrier}, G., {Barman}, T.~S., {Allard}, F., \& {Hauschildt},
  P.~H. 2003, \aap, 402, 701

\bibitem[{{Bayo} {et~al.}(2012){Bayo}, {Barrado}, {Hu{\'e}lamo},
  {Morales-Calder{\'o}n}, {Melo}, {Stauffer}, \& {Stelzer}}]{Bayo2012}
{Bayo}, A., {Barrado}, D., {Hu{\'e}lamo}, N., {et~al.} 2012, \aap, 547, A80

\bibitem[{{Bayo} {et~al.}(2011){Bayo}, {Barrado}, {Stauffer},
  {Morales-Calder{\'o}n}, {Melo}, {Hu{\'e}lamo}, {Bouy}, {Stelzer}, {Tamura},
  \& {Jayawardhana}}]{Bayo2011}
{Bayo}, A., {Barrado}, D., {Stauffer}, J., {et~al.} 2011, \aap, 536, A63

\bibitem[{{Beckwith} {et~al.}(1990){Beckwith}, {Sargent}, {Chini}, \&
  {Guesten}}]{Beckwith1990}
{Beckwith}, S.~V.~W., {Sargent}, A.~I., {Chini}, R.~S., \& {Guesten}, R. 1990,
  \aj, 99, 924

\bibitem[{{Bell} {et~al.}(1997){Bell}, {Cassen}, {Klahr}, \&
  {Henning}}]{Bell1997}
{Bell}, K.~R., {Cassen}, P.~M., {Klahr}, H.~H., \& {Henning}, T. 1997, \apj,
  486, 372

\bibitem[{{Belloche} {et~al.}(2011){Belloche}, {Schuller}, {Parise},
  {Andr{\'e}}, {Hatchell}, {J{\o}rgensen}, {Bontemps}, {Wei{\ss}}, {Menten}, \&
  {Muders}}]{Belloche2011}
{Belloche}, A., {Schuller}, F., {Parise}, B., {et~al.} 2011, \aap, 527, A145

\bibitem[{{Bitner} {et~al.}(2010){Bitner}, {Chen}, {Muzerolle}, {Weinberger},
  {Pecaut}, {Mamajek}, \& {McClure}}]{Bitner2010}
{Bitner}, M.~A., {Chen}, C.~H., {Muzerolle}, J., {et~al.} 2010, \apj, 714, 1542

\bibitem[{{Boehler} {et~al.}(2013){Boehler}, {Dutrey}, {Guilloteau}, \&
  {Pi{\'e}tu}}]{Boehler2013}
{Boehler}, Y., {Dutrey}, A., {Guilloteau}, S., \& {Pi{\'e}tu}, V. 2013, \mnras,
  431, 1573

\bibitem[{{Bouwman} {et~al.}(2008){Bouwman}, {Henning}, {Hillenbrand}, {Meyer},
  {Pascucci}, {Carpenter}, {Hines}, {Kim}, {Silverstone}, {Hollenbach}, \&
  {Wolf}}]{Bouwman2008}
{Bouwman}, J., {Henning}, T., {Hillenbrand}, L.~A., {et~al.} 2008, \apj, 683,
  479

\bibitem[{{Cieza} {et~al.}(2011){Cieza}, {Olofsson}, {Harvey}, {Pinte},
  {Mer{\'{\i}}n}, {Augereau}, {Evans}, {Najita}, {Henning}, \&
  {M{\'e}nard}}]{Cieza2011}
{Cieza}, L.~A., {Olofsson}, J., {Harvey}, P.~M., {et~al.} 2011, \apjl, 741, L25

\bibitem[{{Coupeaud} {et~al.}(2011){Coupeaud}, {Demyk}, {Meny}, {Nayral},
  {Delpech}, {Leroux}, {Depecker}, {Creff}, {Brubach}, \& {Roy}}]{Coupeaud2011}
{Coupeaud}, A., {Demyk}, K., {Meny}, C., {et~al.} 2011, \aap, 535, A124

\bibitem[{{Diolaiti} {et~al.}(2000){Diolaiti}, {Bendinelli}, {Bonaccini},
  {Close}, {Currie}, \& {Parmeggiani}}]{Diolaiti2000}
{Diolaiti}, E., {Bendinelli}, O., {Bonaccini}, D., {et~al.} 2000, in Society of
  Photo-Optical Instrumentation Engineers (SPIE) Conference Series, Vol. 4007,
  Society of Photo-Optical Instrumentation Engineers (SPIE) Conference Series,
  ed. P.~L. {Wizinowich}, 879--888

\bibitem[{{Draine}(2003)}]{Draine2003}
{Draine}, B.~T. 2003, \apj, 598, 1026

\bibitem[{Draine(2006)}]{Draine2006}
Draine, B.~T. 2006, The Astrophysical Journal, 636, 1114

\bibitem[{{Dullemond} \& {Dominik}(2004)}]{Dullemond2004}
{Dullemond}, C.~P. \& {Dominik}, C. 2004, \aap, 417, 159

\bibitem[{{Dullemond} \& {Dominik}(2008)}]{Dullemond2008}
{Dullemond}, C.~P. \& {Dominik}, C. 2008, \aap, 487, 205

\bibitem[{{Fazio} {et~al.}(2004){Fazio}, {Hora}, {Allen}, {Ashby}, {Barmby},
  {Deutsch}, {Huang}, {Kleiner}, {Marengo}, {Megeath}, {Melnick}, {Pahre},
  {Patten}, {Polizotti}, {Smith}, {Taylor}, {Wang}, {Willner}, {Hoffmann},
  {Pipher}, {Forrest}, {McMurty}, {McCreight}, {McKelvey}, {McMurray}, {Koch},
  {Moseley}, {Arendt}, {Mentzell}, {Marx}, {Losch}, {Mayman}, {Eichhorn},
  {Krebs}, {Jhabvala}, {Gezari}, {Fixsen}, {Flores}, {Shakoorzadeh}, {Jungo},
  {Hakun}, {Workman}, {Karpati}, {Kichak}, {Whitley}, {Mann}, {Tollestrup},
  {Eisenhardt}, {Stern}, {Gorjian}, {Bhattacharya}, {Carey}, {Nelson},
  {Glaccum}, {Lacy}, {Lowrance}, {Laine}, {Reach}, {Stauffer}, {Surace},
  {Wilson}, {Wright}, {Hoffman}, {Domingo}, \& {Cohen}}]{Fazio2004}
{Fazio}, G.~G., {Hora}, J.~L., {Allen}, L.~E., {et~al.} 2004, \apjs, 154, 10

\bibitem[{{Furlan} {et~al.}(2009){Furlan}, {Watson}, {McClure}, {Manoj},
  {Espaillat}, {D'Alessio}, {Calvet}, {Kim}, {Sargent}, {Forrest}, \&
  {Hartmann}}]{Furlan2009}
{Furlan}, E., {Watson}, D.~M., {McClure}, M.~K., {et~al.} 2009, \apj, 703, 1964

\bibitem[{{Harvey} {et~al.}(2012{\natexlab{a}}){Harvey}, {Henning}, {Liu},
  {M{\'e}nard}, {Pinte}, {Wolf}, {Cieza}, {Evans}, \& {Pascucci}}]{Harvey2012}
{Harvey}, P.~M., {Henning}, T., {Liu}, Y., {et~al.} 2012{\natexlab{a}}, \apj,
  755, 67

\bibitem[{{Harvey} {et~al.}(2012{\natexlab{b}}){Harvey}, {Henning},
  {M{\'e}nard}, {Wolf}, {Liu}, {Cieza}, {Evans}, {Pascucci}, {Mer{\'{\i}}n}, \&
  {Pinte}}]{Harvey2012a}
{Harvey}, P.~M., {Henning}, T., {M{\'e}nard}, F., {et~al.} 2012{\natexlab{b}},
  \apjl, 744, L1

\bibitem[{{Hauschildt} {et~al.}(1999){Hauschildt}, {Allard}, \&
  {Baron}}]{Hauschildt1999}
{Hauschildt}, P.~H., {Allard}, F., \& {Baron}, E. 1999, \apj, 512, 377

\bibitem[{{Hern{\'a}ndez} {et~al.}(2007){Hern{\'a}ndez}, {Hartmann}, {Megeath},
  {Gutermuth}, {Muzerolle}, {Calvet}, {Vivas}, {Brice{\~n}o}, {Allen},
  {Stauffer}, {Young}, \& {Fazio}}]{Hernandez2007}
{Hern{\'a}ndez}, J., {Hartmann}, L., {Megeath}, T., {et~al.} 2007, \apj, 662,
  1067

\bibitem[{{Houck} {et~al.}(2004){Houck}, {Roellig}, {van Cleve}, {Forrest},
  {Herter}, {Lawrence}, {Matthews}, {Reitsema}, {Soifer}, {Watson}, {Weedman},
  {Huisjen}, {Troeltzsch}, {Barry}, {Bernard-Salas}, {Blacken}, {Brandl},
  {Charmandaris}, {Devost}, {Gull}, {Hall}, {Henderson}, {Higdon}, {Pirger},
  {Schoenwald}, {Sloan}, {Uchida}, {Appleton}, {Armus}, {Burgdorf},
  {Fajardo-Acosta}, {Grillmair}, {Ingalls}, {Morris}, \& {Teplitz}}]{Houck2004}
{Houck}, J.~R., {Roellig}, T.~L., {van Cleve}, J., {et~al.} 2004, \apjs, 154,
  18

\bibitem[{{Joergens} {et~al.}(2012){Joergens}, {Pohl}, {Sicilia-Aguilar}, \&
  {Henning}}]{Joergens2012}
{Joergens}, V., {Pohl}, A., {Sicilia-Aguilar}, A., \& {Henning}, T. 2012, \aap,
  543, A151

\bibitem[{{Kessler-Silacci} {et~al.}(2006){Kessler-Silacci}, {Augereau},
  {Dullemond}, {Geers}, {Lahuis}, {Evans}, {van Dishoeck}, {Blake}, {Boogert},
  {Brown}, {J{\o}rgensen}, {Knez}, \& {Pontoppidan}}]{Kessler-Silacci2006}
{Kessler-Silacci}, J., {Augereau}, J.-C., {Dullemond}, C.~P., {et~al.} 2006,
  \apj, 639, 275

\bibitem[{{Kraus} \& {Hillenbrand}(2007)}]{Kraus2007}
{Kraus}, A.~L. \& {Hillenbrand}, L.~A. 2007, \apj, 662, 413

\bibitem[{{Lafreni{\`e}re} {et~al.}(2008){Lafreni{\`e}re}, {Jayawardhana},
  {Brandeker}, {Ahmic}, \& {van Kerkwijk}}]{Lafreni`ere2008a}
{Lafreni{\`e}re}, D., {Jayawardhana}, R., {Brandeker}, A., {Ahmic}, M., \& {van
  Kerkwijk}, M.~H. 2008, \apj, 683, 844

\bibitem[{{Looper} {et~al.}(2010){Looper}, {Mohanty}, {Bochanski}, {Burgasser},
  {Mamajek}, {Herczeg}, {West}, {Faherty}, {Rayner}, {Pitts}, \&
  {Kirkpatrick}}]{Looper2010}
{Looper}, D.~L., {Mohanty}, S., {Bochanski}, J.~J., {et~al.} 2010, \apj, 714,
  45

\bibitem[{{Luhman}(2007)}]{Luhman2007a}
{Luhman}, K.~L. 2007, \apjs, 173, 104

\bibitem[{{Luhman} {et~al.}(2007){Luhman}, {Adame}, {D'Alessio}, {Calvet},
  {McLeod}, {Bohac}, {Forrest}, {Hartmann}, {Sargent}, \&
  {Watson}}]{Luhman2007b}
{Luhman}, K.~L., {Adame}, L., {D'Alessio}, P., {et~al.} 2007, \apj, 666, 1219

\bibitem[{{Luhman} {et~al.}(2008){Luhman}, {Allen}, {Allen}, {Gutermuth},
  {Hartmann}, {Mamajek}, {Megeath}, {Myers}, \& {Fazio}}]{Luhman2008a}
{Luhman}, K.~L., {Allen}, L.~E., {Allen}, P.~R., {et~al.} 2008, \apj, 675, 1375

\bibitem[{{Luhman} \& {Muench}(2008)}]{Luhman2008}
{Luhman}, K.~L. \& {Muench}, A.~A. 2008, \apj, 684, 654

\bibitem[{{Manoj} {et~al.}(2011){Manoj}, {Kim}, {Furlan}, {McClure}, {Luhman},
  {Watson}, {Espaillat}, {Calvet}, {Najita}, {D'Alessio}, {Adame}, {Sargent},
  {Forrest}, {Bohac}, {Green}, \& {Arnold}}]{Manoj2011}
{Manoj}, P., {Kim}, K.~H., {Furlan}, E., {et~al.} 2011, \apjs, 193, 11

\bibitem[{{Mathews} {et~al.}(2013){Mathews}, {Pinte}, {Duchene}, {Williams}, \&
  {Menard}}]{Mathews2013}
{Mathews}, G.~S., {Pinte}, C., {Duchene}, G., {Williams}, J.~P., \& {Menard},
  F. 2013, ArXiv e-prints

\bibitem[{{Mathis} {et~al.}(1977){Mathis}, {Rumpl}, \&
  {Nordsieck}}]{Mathis1977}
{Mathis}, J.~S., {Rumpl}, W., \& {Nordsieck}, K.~H. 1977, \apj, 217, 425

\bibitem[{{Mordasini} {et~al.}(2012){Mordasini}, {Alibert}, {Benz}, {Klahr}, \&
  {Henning}}]{Mordasini2012}
{Mordasini}, C., {Alibert}, Y., {Benz}, W., {Klahr}, H., \& {Henning}, T. 2012,
  \aap, 541, A97

\bibitem[{{Morrow} {et~al.}(2008){Morrow}, {Luhman}, {Espaillat}, {D'Alessio},
  {Adame}, {Calvet}, {Forrest}, {Sargent}, {Hartmann}, {Watson}, \&
  {Bohac}}]{Morrow2008}
{Morrow}, A.~L., {Luhman}, K.~L., {Espaillat}, C., {et~al.} 2008, \apjl, 676,
  L143

\bibitem[{{Mulders} \& {Dominik}(2012)}]{Mulders2012}
{Mulders}, G.~D. \& {Dominik}, C. 2012, \aap, 539, A9

\bibitem[{{Muzerolle} {et~al.}(2009){Muzerolle}, {Flaherty}, {Balog}, {Furlan},
  {Smith}, {Allen}, {Calvet}, {D'Alessio}, {Megeath}, {Muench}, {Rieke}, \&
  {Sherry}}]{Muzerolle2009}
{Muzerolle}, J., {Flaherty}, K., {Balog}, Z., {et~al.} 2009, \apjl, 704, L15

\bibitem[{{Natta} {et~al.}(2000){Natta}, {Grinin}, \& {Mannings}}]{Natta2000}
{Natta}, A., {Grinin}, V., \& {Mannings}, V. 2000, {Properties and evolution of
  disks around pre-main sequence stars of intermediate mass} ({Protostars and
  Planets IV})

\bibitem[{{Oliveira} {et~al.}(2011){Oliveira}, {Olofsson}, {Pontoppidan}, {van
  Dishoeck}, {Augereau}, \& {Mer{\'{\i}}n}}]{Oliveira2011}
{Oliveira}, I., {Olofsson}, J., {Pontoppidan}, K.~M., {et~al.} 2011, \apj, 734,
  51

\bibitem[{{Olofsson} {et~al.}(2010){Olofsson}, {Augereau}, {van Dishoeck},
  {Mer{\'{\i}}n}, {Grosso}, {M{\'e}nard}, {Blake}, \& {Monin}}]{Olofsson2010}
{Olofsson}, J., {Augereau}, J.-C., {van Dishoeck}, E.~F., {et~al.} 2010, \aap,
  520, A39

\bibitem[{{Olofsson} {et~al.}(2009){Olofsson}, {Augereau}, {van Dishoeck},
  {Mer{\'{\i}}n}, {Lahuis}, {Kessler-Silacci}, {Dullemond}, {Oliveira},
  {Blake}, {Boogert}, {Brown}, {Evans}, {Geers}, {Knez}, {Monin}, \&
  {Pontoppidan}}]{Olofsson2009}
{Olofsson}, J., {Augereau}, J.-C., {van Dishoeck}, E.~F., {et~al.} 2009, \aap,
  507, 327

\bibitem[{{Osterloh} \& {Beckwith}(1995)}]{Osterloh1995}
{Osterloh}, M. \& {Beckwith}, S.~V.~W. 1995, \apj, 439, 288

\bibitem[{{Padoan} \& {Nordlund}(2002)}]{Padoan2002}
{Padoan}, P. \& {Nordlund}, {\AA}. 2002, \apj, 576, 870

\bibitem[{{Pascucci} {et~al.}(2009){Pascucci}, {Apai}, {Luhman}, {Henning},
  {Bouwman}, {Meyer}, {Lahuis}, \& {Natta}}]{Pascucci2009}
{Pascucci}, I., {Apai}, D., {Luhman}, K., {et~al.} 2009, \apj, 696, 143

\bibitem[{{Pilbratt} {et~al.}(2010){Pilbratt}, {Riedinger}, {Passvogel},
  {Crone}, {Doyle}, {Gageur}, {Heras}, {Jewell}, {Metcalfe}, {Ott}, \&
  {Schmidt}}]{Pilbratt2010}
{Pilbratt}, G.~L., {Riedinger}, J.~R., {Passvogel}, T., {et~al.} 2010, \aap,
  518, L1

\bibitem[{{Pinte} {et~al.}(2008){Pinte}, {Padgett}, {M{\'e}nard},
  {Stapelfeldt}, {Schneider}, {Olofsson}, {Pani{\'c}}, {Augereau},
  {Duch{\^e}ne}, {Krist}, {Pontoppidan}, {Perrin}, {Grady}, {Kessler-Silacci},
  {van Dishoeck}, {Lommen}, {Silverstone}, {Hines}, {Wolf}, {Blake}, {Henning},
  \& {Stecklum}}]{Pinte2008}
{Pinte}, C., {Padgett}, D.~L., {M{\'e}nard}, F., {et~al.} 2008, \aap, 489, 633

\bibitem[{{Poglitsch} {et~al.}(2010){Poglitsch}, {Waelkens}, {Geis},
  {Feuchtgruber}, {Vandenbussche}, {Rodriguez}, {Krause}, {Renotte}, {van
  Hoof}, {Saraceno}, {Cepa}, {Kerschbaum}, {Agn{\`e}se}, {Ali}, {Altieri},
  {Andreani}, {Augueres}, {Balog}, {Barl}, {Bauer}, {Belbachir}, {Benedettini},
  {Billot}, {Boulade}, {Bischof}, {Blommaert}, {Callut}, {Cara}, {Cerulli},
  {Cesarsky}, {Contursi}, {Creten}, {De Meester}, {Doublier}, {Doumayrou},
  {Duband}, {Exter}, {Genzel}, {Gillis}, {Gr{\"o}zinger}, {Henning},
  {Herreros}, {Huygen}, {Inguscio}, {Jakob}, {Jamar}, {Jean}, {de Jong},
  {Katterloher}, {Kiss}, {Klaas}, {Lemke}, {Lutz}, {Madden}, {Marquet},
  {Martignac}, {Mazy}, {Merken}, {Montfort}, {Morbidelli}, {M{\"u}ller},
  {Nielbock}, {Okumura}, {Orfei}, {Ottensamer}, {Pezzuto}, {Popesso},
  {Putzeys}, {Regibo}, {Reveret}, {Royer}, {Sauvage}, {Schreiber}, {Stegmaier},
  {Schmitt}, {Schubert}, {Sturm}, {Thiel}, {Tofani}, {Vavrek}, {Wetzstein},
  {Wieprecht}, \& {Wiezorrek}}]{Poglitsch2010}
{Poglitsch}, A., {Waelkens}, C., {Geis}, N., {et~al.} 2010, \aap, 518, L2+

\bibitem[{{Reipurth} \& {Clarke}(2001)}]{Reipurth2001}
{Reipurth}, B. \& {Clarke}, C. 2001, \aj, 122, 432

\bibitem[{{Ricci} {et~al.}(2013){Ricci}, {Isella}, {Carpenter}, \&
  {Testi}}]{Ricci2013}
{Ricci}, L., {Isella}, A., {Carpenter}, J.~M., \& {Testi}, L. 2013, \apjl, 764,
  L27

\bibitem[{{Ricci} {et~al.}(2012){Ricci}, {Testi}, {Natta}, {Scholz}, \& {de
  Gregorio-Monsalvo}}]{Ricci2012}
{Ricci}, L., {Testi}, L., {Natta}, A., {Scholz}, A., \& {de Gregorio-Monsalvo},
  I. 2012, \apjl, 761, L20

\bibitem[{{Rieke} {et~al.}(2004){Rieke}, {Young}, {Engelbracht}, {Kelly},
  {Low}, {Haller}, {Beeman}, {Gordon}, {Stansberry}, {Misselt}, {Cadien},
  {Morrison}, {Rivlis}, {Latter}, {Noriega-Crespo}, {Padgett}, {Stapelfeldt},
  {Hines}, {Egami}, {Muzerolle}, {Alonso-Herrero}, {Blaylock}, {Dole}, {Hinz},
  {Le Floc'h}, {Papovich}, {P{\'e}rez-Gonz{\'a}lez}, {Smith}, {Su}, {Bennett},
  {Frayer}, {Henderson}, {Lu}, {Masci}, {Pesenson}, {Rebull}, {Rho}, {Keene},
  {Stolovy}, {Wachter}, {Wheaton}, {Werner}, \& {Richards}}]{Rieke2004}
{Rieke}, G.~H., {Young}, E.~T., {Engelbracht}, C.~W., {et~al.} 2004, \apjs,
  154, 25

\bibitem[{{Robberto} {et~al.}(2012){Robberto}, {Spina}, {Da Rio}, {Apai},
  {Pascucci}, {Ricci}, {Goddi}, {Testi}, {Palla}, \&
  {Bacciotti}}]{Robberto2012}
{Robberto}, M., {Spina}, L., {Da Rio}, N., {et~al.} 2012, \aj, 144, 83

\bibitem[{{Scholz} {et~al.}(2006){Scholz}, {Jayawardhana}, \&
  {Wood}}]{Scholz2006}
{Scholz}, A., {Jayawardhana}, R., \& {Wood}, K. 2006, \apj, 645, 1498

\bibitem[{{Skrutskie} {et~al.}(2006){Skrutskie}, {Cutri}, {Stiening},
  {Weinberg}, {Schneider}, {Carpenter}, {Beichman}, {Capps}, {Chester},
  {Elias}, {Huchra}, {Liebert}, {Lonsdale}, {Monet}, {Price}, {Seitzer},
  {Jarrett}, {Kirkpatrick}, {Gizis}, {Howard}, {Evans}, {Fowler}, {Fullmer},
  {Hurt}, {Light}, {Kopan}, {Marsh}, {McCallon}, {Tam}, {Van Dyk}, \&
  {Wheelock}}]{Skrutskie2006}
{Skrutskie}, M.~F., {Cutri}, R.~M., {Stiening}, R., {et~al.} 2006, \aj, 131,
  1163

\bibitem[{{Spezzi} {et~al.}(2013){Spezzi}, {Cox}, {Prusti}, {Mer{\'{\i}}n},
  {Ribas}, {Alves de Oliveira}, {Winston}, {K{\'o}sp{\'a}l}, {Royer}, {Vavrek},
  {Andr{\'e}}, {Pilbratt}, {Testi}, {Bressert}, {Ricci}, {Men'shchikov}, \&
  {K{\"o}nyves}}]{Spezzi2013}
{Spezzi}, L., {Cox}, N.~L.~J., {Prusti}, T., {et~al.} 2013, \aap, 555, A71

\bibitem[{{Sz{\H u}cs} {et~al.}(2010){Sz{\H u}cs}, {Apai}, {Pascucci}, \&
  {Dullemond}}]{SzHucs2010}
{Sz{\H u}cs}, L., {Apai}, D., {Pascucci}, I., \& {Dullemond}, C.~P. 2010, \apj,
  720, 1668

\bibitem[{{Werner} {et~al.}(2004){Werner}, {Roellig}, {Low}, {Rieke}, {Rieke},
  {Hoffmann}, {Young}, {Houck}, {Brandl}, {Fazio}, {Hora}, {Gehrz}, {Helou},
  {Soifer}, {Stauffer}, {Keene}, {Eisenhardt}, {Gallagher}, {Gautier}, {Irace},
  {Lawrence}, {Simmons}, {Van Cleve}, {Jura}, {Wright}, \&
  {Cruikshank}}]{Werner2004}
{Werner}, M.~W., {Roellig}, T.~L., {Low}, F.~J., {et~al.} 2004, \apjs, 154, 1

\bibitem[{{Williams} \& {Cieza}(2011)}]{Williams2011}
{Williams}, J.~P. \& {Cieza}, L.~A. 2011, \araa, 49, 67

\bibitem[{{Wright} {et~al.}(2010){Wright}, {Eisenhardt}, {Mainzer}, {Ressler},
  {Cutri}, {Jarrett}, {Kirkpatrick}, {Padgett}, {McMillan}, {Skrutskie},
  {Stanford}, {Cohen}, {Walker}, {Mather}, {Leisawitz}, {Gautier}, {McLean},
  {Benford}, {Lonsdale}, {Blain}, {Mendez}, {Irace}, {Duval}, {Liu}, {Royer},
  {Heinrichsen}, {Howard}, {Shannon}, {Kendall}, {Walsh}, {Larsen}, {Cardon},
  {Schick}, {Schwalm}, {Abid}, {Fabinsky}, {Naes}, \& {Tsai}}]{Wright2010}
{Wright}, E.~L., {Eisenhardt}, P.~R.~M., {Mainzer}, A.~K., {et~al.} 2010, \aj,
  140, 1868

\end{thebibliography}

\end{document}